\documentclass[twocolumn,twocolappendix,trackchanges]{aastex701}
\usepackage{amsmath}	
\usepackage{bm}	        
\usepackage{CJKutf8}    
\usepackage{gensymb}    
\usepackage{physunits}  
\newcommand{\anote}{^{\mathrm{a}}}
\newcommand{\bnote}{^{\mathrm{b}}}
\newcommand{\cnote}{^{\mathrm{c}}}
\newcommand{\dnote}{^{\mathrm{d}}}
\newcommand{\enote}{^{\mathrm{e}}}
\newcommand{\fnote}{^{\mathrm{f}}}
\newcommand{\gnote}{^{\mathrm{g}}}
\newcommand{\hnote}{^{\mathrm{h}}}

\newcommand{\amax}{a_\mathrm{max}}                          
\newcommand{\amin}{a_\mathrm{min}}                          
\newcommand{\csiso}{c_\mathrm{s,iso}}                       
\newcommand{\csison}{c_\mathrm{s,iso,0}}                    
\newcommand{\Er}{E_\mathrm{r}}                              
\newcommand{\Erf}{E_{\mathrm{r},f}}                         
\newcommand{\Ertot}{E_\mathrm{r,tot}}                       
\newcommand{\Frf}{F_{r,f}}                                  
\newcommand{\ftherm}{f_\mathrm{therm}}                      
\newcommand{\firrad}{f_\mathrm{irrad}}                      
\newcommand{\kabs}{\kappa^\mathrm{abs}}                     
\newcommand{\kaf}{\kappa_{\mathrm{a},f}}                    
\newcommand{\kB}{k_\mathrm{B}}                              
\newcommand{\knuabs}{\kappa_\nu^\mathrm{abs}}               
\newcommand{\knuabssca}{\kappa_\nu^\mathrm{abs+sca}}        
\newcommand{\knuabsscag}{\kappa_\nu^{\mathrm{abs+sca}+g}}   
\newcommand{\knusca}{\kappa_\nu^\mathrm{sca}}               
\newcommand{\kpf}{\kappa_{\mathrm{p},f}}                    
\newcommand{\kpfe}{\kappa_{\mathrm{p},f,\mathrm{e}}}        
\newcommand{\krf}{\kappa_{\mathrm{r},f}}                    
\newcommand{\ksf}{\kappa_{\mathrm{s},f}}                    
\newcommand{\kstarabs}{\kappa_*^\mathrm{abs}}               
\newcommand{\kstarsca}{\kappa_*^\mathrm{sca}}               
\newcommand{\nhat}{\hat{\mathbf{{n}}}}                      
\newcommand{\nhatrin}{\nhat_{r,\mathrm{in}}}                
\newcommand{\nhatrout}{\nhat_{r,\mathrm{out}}}              
\newcommand{\Nang}{N_\mathrm{ang}}                          
\newcommand{\nupeak}{\nu_\mathrm{peak}}                     
\newcommand{\od}{\mathrm{d}}                                
\newcommand{\Omegamid}{\Omega_\mathrm{mid}}                 
\newcommand{\Omegapol}{\Omega_\mathrm{pol}}                   
\newcommand{\phimax}{\phi_\mathrm{max}}                     
\newcommand{\phimin}{\phi_\mathrm{min}}                     
\newcommand{\rhodtot}{\rho_\mathrm{d,tot}}                  
\newcommand{\rhomin}{\rho_\mathrm{min}}                     
\newcommand{\rhosig}{\rho_\mathrm{sig}}                     
\newcommand{\rmax}{r_\mathrm{max}}                          
\newcommand{\rmin}{r_\mathrm{min}}                          
\newcommand{\Tc}{T_\mathrm{c}}                              
\newcommand{\Ttherm}{T_\mathrm{therm}}                      
\newcommand{\tcross}{t_\mathrm{cross}}                      
\newcommand{\tlim}{t_\mathrm{lim}}                          
\newcommand{\thetamid}{\theta_\mathrm{mid}}                 
\newcommand{\Teq}{T_\mathrm{eq}}                            
\newcommand{\teq}{t_\mathrm{eq}}                            
\newcommand{\Tmax}{T_\mathrm{max}}                          
\newcommand{\Tmin}{T_\mathrm{min}}                          
\newcommand{\thetamax}{\theta_\mathrm{max}}                 
\newcommand{\thetamin}{\theta_\mathrm{min}}                 
\newcommand{\zetaf}{\zeta^\mathrm{f}}                       

\begin{document}
\begin{CJK*}{UTF8}{gbsn}
\title{A Framework to Model Stellar Irradiated Disks with Frequency-dependent Absorption and Scattering Opacities in Athena++}

\correspondingauthor{Stanley A. Baronett}
\author[orcid=0000-0003-0412-760X,gname=Stanley,sname=Baronett]{Stanley A. Baronett}
\altaffiliation{UNLV Foundation Board of Trustees Graduate Fellow}
\affiliation{Center for Computational Astrophysics, Flatiron Institute, 162 5th Ave, New York, NY 10010, USA}
\affiliation{Nevada Center for Astrophysics, University of Nevada, Las Vegas, Box 454002, Las Vegas, NV 89154, USA}
\affiliation{Department of Physics and Astronomy, University of Nevada, Las Vegas (UNLV), Box 454002, Las Vegas, NV 89154, USA}
\email[show]{barons2@unlv.nevada.edu}

\author[0000-0002-2624-3399]{Yan-Fei Jiang (姜燕飞)}
\affiliation{Center for Computational Astrophysics, Flatiron Institute, 162 5th Ave, New York, NY 10010, USA}
\email{yjiang@flatironinstitute.org}

\author[0000-0003-3616-6822]{Zhaohuan Zhu (朱照寰)}
\affiliation{Nevada Center for Astrophysics, University of Nevada, Las Vegas, Box 454002, Las Vegas, NV 89154, USA}
\affiliation{Department of Physics and Astronomy, University of Nevada, Las Vegas, Box 454002, Las Vegas, NV 89154, USA}
\email{zhaohuan.zhu@unlv.edu}

\author[0000-0002-8537-9114]{Shangjia Zhang (张尚嘉)}
\affiliation{
Department of Astronomy, Columbia University, Mail Code 5246, New York, NY 10027, USA}
\affiliation{Nevada Center for Astrophysics, University of Nevada, Las Vegas, Box 454002, Las Vegas, NV 89154, USA}
\affiliation{Department of Physics and Astronomy, University of Nevada, Las Vegas, Box 454002, Las Vegas, NV 89154, USA}
\email{sz3342@columbia.edu}

\author[0000-0001-5032-1396]{Philip J. Armitage}
\affiliation{Center for Computational Astrophysics, Flatiron Institute, 162 5th Ave, New York, NY 10010, USA}
\affiliation{Department of Physics and Astronomy, Stony Brook University, Stony Brook, NY 11794, USA}
\email{parmitage@flatironinstitute.org}

\begin{abstract}

The frequency dependence of opacity is crucial for determining the thermal structure of protoplanetary disks, which in turn influences disk dynamics and planet formation.
Yet many disk models adopt simplified thermodynamics, and common radiation-hydrodynamic approaches often use gray opacities, ignore scattering, and yield inaccurate results in regions with intermediate optical depth.
We present a comprehensive framework that models stellar irradiation with frequency-dependent absorption and scattering across all optical depths using the Athena++ finite-volume code, extended with multigroup radiation transport and newly implemented radial rays to more accurately represent the stellar flux.
To calibrate this framework, we focus exclusively on hydrostatic disk models, allowing us to isolate radiative effects and evaluate the method without additional dynamical complexity.
Because dust opacity increases strongly with frequency, ultraviolet stellar irradiation heats the tenuous disk atmosphere while the optically thick midplane remains cooler.
This vertical temperature gradient is captured more accurately when more frequency bands are used or when scattering is included.
Our hydrostatic models achieve equilibrium temperatures that differ from Monte Carlo radiative-transfer benchmarks on average by 2--5\% with 64 frequency bands and 7--11\% with 3 bands.
Reducing the number of bands lowers computational cost by at least an order of magnitude while increasing the maximum possible temperature deviation only from 8\% to 19\%.
This calibration demonstrates the accuracy and efficiency of the framework and provides a solid foundation for future self-consistent studies of irradiated protoplanetary disks, including fully dynamical simulations and applications involving chemical processes and time-dependent stellar luminosity.

\end{abstract}



\section{Introduction} \label{sec:introduction}

Stellar irradiation is the primary heating mechanism that predominantly sets the temperature throughout a protoplanetary disk \citep{D'AlessioCantoCalvet1998}.
This includes complex vertical thermal structures \citep{LawTeagueLoomis2021, Galloway-SprietsmaBaeIzquierdo2025} recently revealed by observations of molecular-line emission from a variety of sources \citep{ObergGuzmanWalsh2021, TeagueBenistyFacchini2025}.
Radiation and the resultant structure directly affect complex hydrodynamics, including thermodynamically sensitive instabilities driven by self-gravity, baroclinicity, and buoyancy effects (see \citealt{LyraUmurhan2019} and \citealt{LesurFlockErcolano2023} for recent reviews).
Radiation can also influence various aspects of planet formation and migration (see \citealt{DrazkowskaBitschLambrechts2023} and \citealt{PaardekooperDongDuffell2023} for recent reviews).

A significant amount of stellar spectral energy is absorbed by dust, whose spatial distribution results from a balance between sedimentation and turbulence \citep{DullemondDominik2004}.
Meanwhile, as the grain size distribution evolves \citep{DullemondDominik2005, BirnstielKlahrErcolano2012}, so does its overall frequency-dependent opacity \citep{BirnstielDullemondZhu2018}.
Together, these affect the temperature structure \citep{ChiangGoldreich1997, D'AlessioCantoCalvet1998, SenguptaDodson-RobinsonHasegawa2019} and chemical evolution of disks, which involves numerous sublimation regions that result in a complex radial distribution of volatiles \citep{LawLoomisTeague2021} whose condensation and subsequent photochemistry occur on grain surfaces \citep{ObergBergin2021}, which also affects opacities.

Theoretical and computational models require radiation transport to properly account for heating and cooling processes.
Compared to simplified moment equations, the complete angular and frequency dependence of radiative transfer strongly influences the vertical temperature structure \citep{DullemondvanZadelhoffNatta2002}.
By individually tracking an immense number of photon packets and their interactions through a static medium, Monte Carlo methods have accurately reproduced the spectral energy distributions, dust continuum emission, and scattered light observations of many disks.
These achievements have established them as reliable benchmarks against alternative methods \citep{PascucciWolfSteinacker2004, PinteFouchetMenard2007}.

Unfortunately, combining Monte Carlo radiative transfer with hydrodynamics in a self-consistent way poses major conceptual and computational challenges.
Although the review by \citealt{NoebauerSim2019}) summarizes several approaches for mitigating the numerically stiff energy exchange produced by radiative interaction terms---stiffness that likewise limits deterministic (i.e., non--Monte Carlo) transport schemes---practically none of the Monte Carlo codes used for low-energy or nonrelativistic applications is readily compatible with hydrodynamics \citep[e.g.,][]{MelonFuksmanKlahr2022}.
As a result, other radiation-hydrodynamic methods have provided more effective attempts to investigate self-consistent radiative effects on disk dynamics \citep{MuleyMelonFuksmanKlahr2024}, disk instabilities \citep{ZhangZhuJiang2024}, and even circumplanetary envelopes \citep{KrappKratterYoudin2024, BaileyStoneFung2024}.
However, these methods often trade accuracy for performance.

Flux-limited diffusion \citep{LevermorePomraning1981} has long been and is still widely used in radiation-hydrodynamic modeling \citep{Kley1989, BodenheimerYorkeRozyczka1990, KlahrHenningKley1999, TurnerStone2001, KlahrKley2006, KrumholzKleinMcKee2007, vanderHolstTothSokolov2011, ZhangHowellAlmgren2011, KuiperYorkeMignone2020, KriegerKlahrMelonFuksman2025}.
In this radiative transfer approximation, the typically gray (i.e., frequency-independent) flux follows either a free-streaming or diffusion limit in optically thin or thick regions, respectively.
Consequently, it is well known to give inaccurate solutions where optical depths transition from thin to thick \citep{BoleyDurisenNordlund2007, KrumholzThompson2013}, e.g., the irradiated surface between the disk atmosphere and midplane.
Moreover, due to excessive diffusivity in the free-streaming limit, it is unable to cast shadows \citep{HayesNorman2003}, which may be important for disks that are flared, warped, misaligned, or contain circumplanetary disks or large-scale substructures \citep{CuroneFacchiniAndrews2025}.

The M1 closure \citep{Levermore1984} is another common approach to radiation hydrodynamics with an advantage over flux-limited diffusion in its ability to cast shadows \citep[][\S~3.3 and Fig.~5]{WeihOlivaresRezzolla2020}.
However, since both moment-based methods average over all solid angles in the radiative intensity, they inadvertently merge intersecting beams (\citealt{WeihOlivaresRezzolla2020}, Fig.~3; \citealt{MelonFuksmanFlockKlahr2025}, Fig.~2) instead of allowing them to cross (cf. \citealt{DavisStoneJiang2012}, Fig.~6; \citealt{JiangStoneDavis2014}, Fig.~6; \citealt{Jiang2021}, Fig.~3).
These artificial interactions can reduce the cooling efficiency of the disk, as the outward thermal emission from the midplane encounters the inward emission from the irradiated upper layers \citep{MelonFuksmanKlahr2022}.

Unlike flux-limited diffusion or M1, the discrete ordinate method avoids these limitations by solving the complete transfer equation without assuming closure relations \citep[][\S~3.2]{DavisStoneJiang2012}.
The time-dependent equation can be solved implicitly \citep{Jiang2021} to avoid explicit time-step constraints by the speed of light, e.g., in the nonrelativistic environments of protoplanetary disks.
The method can also be extended by a multigroup approach \citep{Jiang2022} to capture the inherent frequency dependence of radiation and opacities, including scattering.
Its accuracy and advantages, as implemented in the open-source astrophysical magnetohydrodynamic code Athena++ \citep{StoneTomidaWhite2020}, have led \cite{MaPakmorJustham2025} to recently add it to the magnetohydrodynamic code AREPO \citep{Springel2010, PakmorBauerSpringel2011, WeinbergerSpringelPakmor2020}.

Extending the multigroup discrete‑ordinates implementation of \citep{Jiang2022}, we develop an improved framework that incorporates frequency-dependent dust opacities,  and introduces radially aligned rays to better resolve radial transport, enabling more accurate modeling of stellar-irradiated disks in Athena++ (with resources available in a public repository; Appendix~\ref{appx:public_repository}).
As a convenient reference, Table~\ref{tab:mathematical_notation} consolidates the relevant mathematical notation.
Section~\ref{sec:methodology} reviews the nonrelativistic multigroup radiation transport module for Athena++ and introduces our modifications and framework.
Section~\ref{sec:model} specifies an irradiated disk model and dust opacities to demonstrate our framework.
Section~\ref{sec:hydrostatic_comparisons} evaluates its accuracy, convergence, and performance by comparing hydrostatic models with Monte Carlo radiative transfer methods.
Section~\ref{sec:discussion} discusses alternative approaches to radiation hydrodynamics and potential applications of our framework.
Section~\ref{sec:conclusions} concludes with a summary.

\begin{deluxetable*}{llcll}
    \tablecaption{Mathematical Notation\label{tab:mathematical_notation}}
    \tablecolumns{5}
    \tablehead{
        \colhead{Notation}      & \colhead{Definition}                                                  &\colhead{Section}                              &\colhead{Equations}                                                                                        & \colhead{References}      \\
        \multicolumn{1}{c}{(1)} & \multicolumn{1}{c}{(2)}                                               & \multicolumn{1}{c}{(3)}                       & \multicolumn{1}{c}{(4)}                                                                                   & \multicolumn{1}{c}{(5)}}
    \startdata  
        $B_\nu(T)$              & Planck's law for specific intensity                                   & \ref{sec:stellar_irradiation}                 & \eqref{eq:B_nu}, \eqref{eq:kpfe}, \eqref{eq:ksf}, \eqref{eq:krf}                                          & (1, eq.~1.51)             \\
        $\Er$                   & Integrated $\Erf$                                                     & \ref{sec:thermal_equilibrium}                 & \eqref{eq:Er}, \eqref{eq:Ertot}                                                                           & (2, eq.~8)                \\
        $\Erf$                  & Radiation energy density                                              & \ref{sec:irradiative--thermal_equilibrium}    & \eqref{eq:Erf}, \eqref{eq:Er}                                                                             & (2, eq.~7)                \\
        $\Ertot$                & Domain-total $\Er$                                                    & \ref{sec:thermal_equilibrium}                 & \eqref{eq:Ertot}, \eqref{eq:Er}                                                                           & \nodata                   \\
        $f$                     & Frequency band $[\nu_f, \nu_{f+1})$                                   & \ref{sec:radiation_transport}                 & \nodata                                                                                                   & (2, \S~2.1)               \\
        $f_*$                   & $f$ where $\nu_f < \nu_* < \nu_{f+1}$                                 & \ref{sec:dust_opacities}                      & \eqref{eq:f_*-to-nu_*_limit}                                                                              & \nodata                   \\
        $F_r$                   & Integrated $\Frf$                                                     & \ref{sec:thermal_equilibrium}                 & \eqref{eq:Fr}, \eqref{eq:F_r-tau}                                                                         & (2, eq.~8), (1, eq.~1.5a) \\
        $\Frf$                  & Net radial flux                                                       & \ref{sec:thermal_equilibrium}                 & \eqref{eq:Frf}, \eqref{eq:Fr}                                                                             & (2, eq.~7), (1, eq.~1.3b) \\
        $I_{0,f}$               & Lorentz-transformed $I_f$ (comoving frame)                            & \ref{sec:emissivity_opacity}                  & \eqref{eq:comoving_sources}                                                                               & (2, eq.~5)                \\
        $I_f$                   & Band-specific intensity (lab frame)                                   & \ref{sec:radiation_transport}                 & \eqref{eq:radiation_transport}, \eqref{eq:I_f-rout}, \eqref{eq:Erf}, \eqref{eq:Frf}                       & (2, eq.~1)                \\
        $J_{0,f}$               & Mean intensity (comoving frame)                                       & \ref{sec:emissivity_opacity}                  & \eqref{eq:comoving_sources}                                                                               & (2, eq.~13), (1, eq.~1.8) \\
        $\nhat$                 & Cell-centered angles (directions)                                     & \ref{sec:radiation_transport}                 & \eqref{eq:radiation_transport}, \eqref{eq:Erf}, \eqref{eq:Frf}                                            & (3, \S~3.1), (2, \S~2.2)  \\
        $\Nang$                 & Number of $\nhat$ per cell                                            & \ref{sec:radiation_transport}                 & \eqref{eq:w_n-to-dOmega_limit}, \eqref{eq:Erf}, \eqref{eq:Frf}                                            & (3, \S~3.1), (2, \S~2.2)  \\
        $N_f$                   & Number of $f$                                                         & \ref{sec:radiation_transport}                 & \eqref{eq:kf-to-knu_limit}, \eqref{eq:f_*-to-nu_*_limit},\eqref{eq:Er}, \eqref{eq:Fr}                     & (2, \S~2.1)               \\
        $\nhat_r$               & $\nhat$ aligned to cell-centered poles (global $\hat{\mathbf{{r}}}$)  & \ref{sec:stellar_irradiation}                 & \nodata                                                                                                   & \nodata                   \\
        $\nhatrout$             & Radially outward $\nhat_r$                                            & \ref{sec:stellar_irradiation}                 & \eqref{eq:I_f-rout}, \eqref{eq:I-rout}                                                                    & \nodata                   \\
        $N_\zeta$               & Number of cell-centered polar angles $(0, \pi/2)$                     & \ref{sec:radiation_transport}                 & \nodata                                                                                                   & (3, \S~3.2.4)             \\
        $N_\psi$                & Number of cell-centered longitudinal angles $(0, \pi)$                & \ref{sec:radiation_transport}                 & \nodata                                                                                                   & (3, \S~3.2.4)             \\
        $R_*$                   & Stellar (photosphere) radius                                          & \ref{sec:stellar_irradiation}                 & \eqref{eq:I_f-rout}, \eqref{eq:I-rout}, \eqref{eq:F_r-tau}                                                & \nodata                   \\
        $T_*$                   & Effective stellar temperature                                         & \ref{sec:stellar_irradiation}                 & \eqref{eq:I_f-rout}, \eqref{eq:I-rout}, \eqref{eq:F_r-tau}                                                & \nodata                   \\
        $\Tc$                   & Color temperature                                                     & \ref{sec:irradiative--thermal_equilibrium}    & \nodata                                                                                                   & (1, p.~26)                \\
        $\tcross$               & Light-crossing time (e.g., for domain)                                & \ref{sec:thermal_equilibrium}                 & \nodata                                                                                                   & \nodata                   \\
        $\Teq$                  & Equilibrium (dust) temperature                                        & \ref{sec:thermal_equilibrium}                 & \nodata                                                                                                   & \nodata                   \\
        $u_\nu(T)$              & Planck's law for spectral energy density [$4\pi B_\nu(T)/c$]          & \ref{sec:emissivity_opacity}                  & \eqref{eq:thermal_emissivity}                                                                             & (1, eqs.~1.7, 1.8)        \\
        $w_n$                   & Angular quadrature weight of $\nhat$ ($1/\Nang$)                      & \ref{sec:radiation_transport}                 & \eqref{eq:w_n-to-dOmega_limit}, \eqref{eq:I_f-rout}, \eqref{eq:I-rout}, \eqref{eq:Erf}, \eqref{eq:Frf}    & (3, eq.~11), (2, \S~2.2)  \\
        $\varepsilon_{0,f}$     & Thermal emissivity (comoving frame)                                   & \ref{sec:emissivity_opacity}                  & \eqref{eq:thermal_emissivity}, \eqref{eq:comoving_sources}                                                & (2, eq.~12)               \\
        $\eta_f$                & Emissivity (lab frame)                                                & \ref{sec:radiation_transport}                 & \eqref{eq:radiation_transport}                                                                            & (2, \S~2.3)               \\
        $\kstarabs$             & $\knuabs$ at $\nu_*$                                                  & \ref{sec:stellar-irradiated_disk}             & \eqref{eq:tau_*}                                                                                          & \nodata                   \\
        $\kstarsca$             & $\knusca$ at $\nu_*$                                                  & \ref{sec:stellar-irradiated_disk}             & \eqref{eq:tau_*}                                                                                          & \nodata                   \\
        $\kaf$                  & Rosseland mean absorption opacity                                     & \ref{sec:emissivity_opacity}                  & \eqref{eq:kaf}, \eqref{eq:comoving_sources}, \eqref{eq:kf-to-knu_limit}                                   & (2, eq.~14)               \\
        $\kpf$                  & Planck mean absorption opacity ($\varepsilon_{0,f}$ coefficient)      & \ref{sec:emissivity_opacity}                  & \eqref{eq:kpfe}, \eqref{eq:comoving_sources}, \eqref{eq:kf-to-knu_limit}                                  & (2, eq.~14)               \\
        $\kpfe$                 & Planck mean absorption opacity ($J_{0,f}$ coefficient)                & \ref{sec:emissivity_opacity}                  & \eqref{eq:kpfe}, \eqref{eq:comoving_sources}, \eqref{eq:kf-to-knu_limit}                                  & (2, eq.~14)               \\
        $\krf$                  & Rosseland mean opacity                                                & \ref{sec:emissivity_opacity}                  & \eqref{eq:krf}                                                                                            & (1, eq.~1.110)            \\
        $\ksf$                  & Pure-scattering mean opacity                                          & \ref{sec:emissivity_opacity}                  & \eqref{eq:ksf}, \eqref{eq:comoving_sources}, \eqref{eq:kf-to-knu_limit}                                   & (2, \S~2.3)               \\
        $\knuabs$               & Monochromatic absorption opacity                                      & \ref{sec:emissivity_opacity}                  & \eqref{eq:kpfe}, \eqref{eq:krf}, \eqref{eq:kf-to-knu_limit}                                               & (2, \S~2.3)               \\
        $\knusca$               & Monochromatic scattering opacity                                      & \ref{sec:emissivity_opacity}                  & \eqref{eq:ksf}, \eqref{eq:krf}, \eqref{eq:kf-to-knu_limit}                                                & \nodata                   \\
        $\nu_*$                 & $\nupeak$ for $T_*$                                                   & \ref{sec:stellar-irradiated_disk}             & \eqref{eq:f_*-to-nu_*_limit}                                                                              & \nodata                   \\
        $\nupeak$               & Frequency at $\max[B_\nu(T)]$                                         & \ref{sec:irradiative--thermal_equilibrium}    & \eqref{eq:nupeak}                                                                                         & (1, eq.~1.56a)            \\
        $\tau_*$                & Effective optical depth at $\nu_*$                                    & \ref{sec:stellar-irradiated_disk}             & \eqref{eq:tau_*}                                                                                          & (1, eqs.~1.26, 1.98)      \\
        $\chi_f$                & Opacity (lab frame)                                                   & \ref{sec:radiation_transport}                 & \eqref{eq:radiation_transport}                                                                            & (2, \S~2.3)               \\
    \enddata
    \tablecomments{Columns give the (1) notation, (2) definition, (3) section where introduced, (4) equations where defined or used, and (5) references as sources or for comparison.}
    \textbf{References.} (1) \cite{RybickiLightman1979}, (2) \cite{Jiang2022}, (3) \cite{Jiang2021}.
\end{deluxetable*}

\section{Methodology} \label{sec:methodology}

This section introduces the numerical methods and framework for implementing the stellar irradiation and frequency-dependent opacities of protoplanetary disks in Athena++.
Written in C++ as a successor to Athena \citep{StoneGardinerTeuben2008}, the fluid equations are solved by the finite-volume method with native support for mesh refinement which can statically or dynamically enhance spatial resolution only where desired.
Domain decomposition can also be parallelized for strong scaling\footnote{
Strong scaling (cf. weak scaling) entails a reduction in the elapsed real time for execution by increasing the number of parallel processes for a fixed problem size.\label{foot:strong_scaling}}
on high-performance-computing resources.
Moreover, the code is modular and supports various physics, including multigroup radiation transport via discrete ordinates \citep{Jiang2022}, which can be solved explicitly or implicitly \citep{Jiang2021}.

We divide the methodology into the following:
Section~\ref{sec:radiation_transport} summarizes the aspects of the radiation transport module most relevant to the overall framework;
Section~\ref{sec:stellar_irradiation} introduces the enhancement we have made to the module and how we use it to model stellar irradiation;
Section~\ref{sec:emissivity_opacity} details how the emissivity and opacities are computed;
and Section~\ref{sec:irradiative--thermal_equilibrium} explains how to better use temperatures to determine mean opacities in a multigroup framework.

\subsection{Radiation Transport} \label{sec:radiation_transport}

The module supports both frequency-independent and -dependent radiation transport.
Under the multigroup approach \citep{MihalasMihalas1984, VaytetAuditDubroca2011, ZhangHowellAlmgren2013}, the frequency domain $\nu$ can be divided into $N_f$ bands, where $f = 0, 1, ..., N_f - 1 = [0, \nu_1), [\nu_1, \nu_2), ..., [\nu_{N_f - 1}, \infty)$.
The extended module solves for the frequency-integrated specific intensities $I_f \equiv \int_{\nu_f}^{\nu_{f+1}}I_\nu\od\nu$ at time $t$ in the lab frame via the time-dependent radiative transfer equation \citep[cf.][eq.~10]{Jiang2022},
\begin{equation}
    \frac{1}{c}\frac{\partial I_f}{\partial t} + \nhat\bm{\cdot\nabla} I_f = \eta_f - \chi_f I_f,
    \label{eq:radiation_transport}
\end{equation}
where emissivity $\eta_f \equiv \int_{\nu_f}^{\nu_{f+1}}\eta_\nu\od\nu$ and opacity $\chi_f \equiv \int_{\nu_f}^{\nu_{f+1}}\chi_\nu I_\nu\od\nu/I_f$.
By solving equation~\eqref{eq:radiation_transport} implicitly \citep[][\S~3.2.2]{Jiang2021},\footnote{
This can be enabled by configuring (and compiling) Athena++ with the \texttt{-implicit\_radiation} flag.
However, using the \texttt{-nr\_radiation} flag instead solves equation~\eqref{eq:radiation_transport} explicitly where the speed of light limits the time step.\label{foot:implicit_flag}}
the hydrodynamic time step is not limited by the light-crossing time for each cell due to the speed of light $c$.

\begin{figure}
    \includegraphics[width=\columnwidth]{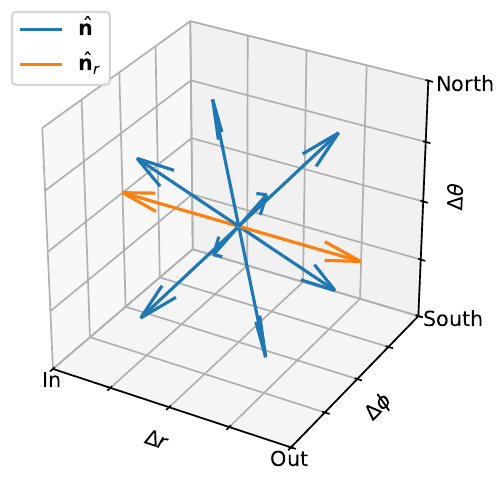}
    \caption{Visualizing the angular discretization within a finite-volume cell with global spherical coordinates $(r, \theta, \phi)$.
    Arrows show the directional unit vectors $\nhat$ used in equation~\eqref{eq:radiation_transport}.
    Blue ones show the standard set described by \citet[][\S~3.2.4]{Jiang2021} for $N_\zeta = 1$ and $N_\psi = 2$ (Section~\ref{sec:radiation_transport}).
    Orange ones show the new pair $\nhat_r$ aligned to the global radial direction $\hat{r}$ (Section~\ref{sec:stellar_irradiation}).}
    \label{fig:nhat}
\end{figure}

The module also solves for $I_f$ along a discrete set of unit vectors $\nhat$ within each finite-volume cell.
If Athena++ is configured to use (global) spherical coordinates $(r, \theta, \phi)$ and \texttt{angle\_flag = 1} is set under the \texttt{<radiation>} block in the input file (Appendix~\ref{appx:public_repository}), the directions of $\nhat$ remain fixed with respect to a local spherical coordinate system centered on each cell whose poles are always aligned to the global radial direction $\hat{r}$ (orange arrows in Figure~\ref{fig:nhat}).
Their angular discretization is described by \citet[][\S~3.2.4]{Jiang2021}, where $N_\zeta \in \mathbb{N}^+$ defines the number of local polar angles between 0 and $\pi/2$, and $N_\psi \in \mathbb{N}^+$
defines the number of local longitudinal angles between 0 and $\pi$ ($N_\psi \geq 2$),\footnote{
In the code, \texttt{npsi = 2} must be the minimum to evenly cover the local hemisphere (cf. Figure~\ref{fig:nhat}).}
the latter of which does not change the solution for axisymmetric models.
As the local quadrants mirror each other, the total number of $\nhat$ angles per cell $\Nang = 4N_\zeta N_\psi$ by default (cf. Section~\ref{sec:stellar_irradiation} and Appendix~\ref{appx:modified_angular_discretization}), and
\begin{equation}
    \lim_{\Nang\to\infty} 4\pi\sum_{n=0}^{\Nang-1}w_n = \int\od\Omega,
    \label{eq:w_n-to-dOmega_limit}
\end{equation}
where $w_n = 1/\Nang \simeq \od\Omega/4\pi$ is the angular quadrature weight of each $\nhat$ angle \citep[cf.][\S~3.1 and eq.~11]{Jiang2021}.
Figure~\ref{fig:nhat} visualizes standard $\nhat$ angles as blue arrows for $N_\zeta = 1$ and $N_\psi = 2$.
Except for the angular resolution study in Section~\ref{sec:angles}, we use $N_\zeta = 4$ and $N_\psi = 2$ in our subsequent models (Section~\ref{sec:model}).

\subsection{Stellar Irradiation} \label{sec:stellar_irradiation}

To model stellar irradiation at distances where the central star can be treated as a point source, we have added two new angles to the radiation transport module (Section~\ref{sec:radiation_transport}) that point inward $\nhatrin$ and outward $\nhatrout$ along the global radial direction.
As detailed in Appendix~\ref{appx:modified_angular_discretization}, these additional $\nhat_r$ increase $\Nang$ and require slight modifications to the default algorithm adopted for angular discretization to ensure equal weights $w_n$.
Since these $\nhat_r$ are aligned to the poles of the local spherical coordinate system centered at each cell, they can be enabled in the latest public versions of Athena++ by setting the optional parameter \texttt{polar\_angle = 1} under the \texttt{<radiation>} block in the input file (Appendix~\ref{appx:public_repository}).
Figure~\ref{fig:nhat} visualizes these new $\nhat_r$ as orange arrows.

Our Athena++ problem generator (Appendix~\ref{appx:public_repository}) allows stellar flux to enter the computational domain only along $\nhatrout$, assuming the inner radial boundary $\rmin$ is sufficiently distant from the star centered at the origin with photosphere radius $R_*$ (i.e., $\rmin \gg R_*$).
Based on the expected flux [as substituted into equation~\eqref{eq:Frf}], the specific intensity \citep[cf.][eqs.~1.3b and 1.13]{RybickiLightman1979} 
\begin{equation}
    I_f(r < \rmin, \nhatrout) = \frac{\pi}{4\pi w_n}\left(\frac{R_*}{r}\right)^2\int_{\nu_f}^{\nu_{f+1}}B_\nu(T_*)\od\nu
    \label{eq:I_f-rout}
\end{equation}
is set at every ghost cell radially inward of $\rmin$, where the effective stellar temperature $T_*$ is used in Planck's law for specific intensity
\begin{equation}
    B_\nu(T) = \frac{2h\nu^3/c^2}{\exp(h\nu/\kB T) - 1},
    \label{eq:B_nu}
\end{equation}
with Planck and Boltzmann constants $h$ and $\kB$, respectively.
When $N_f = 1$ for a frequency-independent approximation, where $f = 0$ and $[\nu_0, \nu_1) = [0, \infty)$ only (Section~\ref{sec:radiation_transport}), equation~\eqref{eq:I_f-rout} simplifies to the Stefan--Boltzmann law, such that \citep[cf.][eq.~1.43]{RybickiLightman1979}
\begin{equation}
    I(r < \rmin, \nhatrout) = \frac{\sigma T_*^4}{4\pi w_n}\left(\frac{R_*}{r}\right)^2,
    \label{eq:I-rout}
\end{equation}
where $\sigma$ is the Stefan--Boltzmann constant.
Although the current problem generator implements a time-constant stellar flux, equations~\eqref{eq:I_f-rout} or \eqref{eq:I-rout} can also depend on time as discussed in Section~\ref{sec:applications}.

\subsection{Emissivity and Opacity} \label{sec:emissivity_opacity}

To determine the emissivity $\eta_f$ and opacity $\chi_f$ source terms in the lab frame [equation~\eqref{eq:radiation_transport}], the module (Section~\ref{sec:radiation_transport}) first computes the source terms in the comoving frame before transforming back, accounting for the proper Doppler shifts in frequencies between frames.
After Lorentz transformation \citep[][eq.~2]{Jiang2021} and integration over the frequency band $[\nu_f, \nu_{f+1})$, the transfer equation in the comoving frame has the source terms \citep[cf.][eq.~11]{Jiang2022}
\begin{align*}
    &\rho\ksf(J_{0,f} - I_{0,f}) + \rho\kaf(\varepsilon_{0,f} - I_{0,f}) \\
    &+ \rho(\kpf - \kaf)(\varepsilon_{0,f} - J_{0,f}) \\
    =\,&\rho(\ksf + \kaf)(J_{0,f} - I_{0,f}) \\
    &+ \rho(\kpf\varepsilon_{0,f} - \kpfe J_{0,f})\stepcounter{equation}\tag{\theequation}\label{eq:comoving_sources}\\
\end{align*}
with local fluid density $\rho$, where the specific intensity $I_{0,f}$, mean intensity $J_{0,f} \equiv (1/4\pi)\int I_{0,f}\od\Omega_0$ (cf. eq.~1.8 of \citealt{RybickiLightman1979} and eq.~8 of \citealt{Jiang2021}), and thermal emissivity $\varepsilon_{0,f}$ are all defined in the comoving frame.
The module allows users to provide the thermal emissivity but by default assumes local thermodynamic equilibrium, where the local temperature $T$ determines \citep[cf.][eq.~12]{Jiang2022}\footnote{
We derive our definition of thermal emissivity $\varepsilon_{0,f}$ from \cite{RybickiLightman1979}, which differs from \citet[][eq.~12]{Jiang2022} by a factor of $c$ due to inherently different definitions (and physical dimensions) of intensity $I$.}
\begin{equation}
   \varepsilon_{0,f} \equiv \frac{c}{4\pi}\int_{\nu_f}^{\nu_{f+1}}u_\nu(T)\od\nu,
    \label{eq:thermal_emissivity}
\end{equation}
where the spectral energy density according to Planck's law is $u_\nu(T) = 4\pi B_\nu(T)/c$ \citep[cf.][eqs.~1.7 and 1.8]{RybickiLightman1979}.\footnote{
Although $u_\nu = (1/c)\int I_\nu\od\Omega$ \citep[][p.~6]{RybickiLightman1979}, the module assumes $u_f \equiv \Erf$ [equation~\eqref{eq:Erf}], which implies $I_f = I_\nu/c$.
However, the module adopts a unit system where factors of $1/c$ or $1/4\pi$ [e.g., when modeling equations~\eqref{eq:I_f-rout} and \eqref{eq:I-rout} in the problem generator (Appendix~\ref{appx:public_repository})] are not needed.}
Absorption, (re)emission, and scattering are assumed to be isotropic in the comoving frame.

\begin{figure*}
    \includegraphics[width=\textwidth]{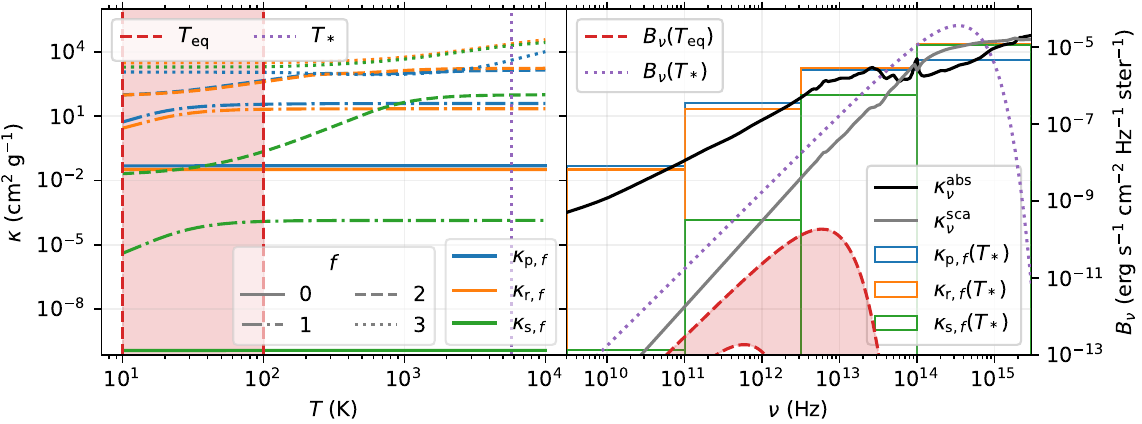}
    \caption{Example dust opacities and relevant Planck spectra.
    In the left panel, curve colors show the Planck mean $\kpf$ [equation~\eqref{eq:kpfe}], Rosseland mean $\krf$ [equation~\eqref{eq:krf}], and pure-scattering-mean $\ksf$ [equation~\eqref{eq:ksf}] opacities as functions of temperature for each one of $N_f = 4$ log-uniform frequency bands $f$ differentiated by line styles (Section~\ref{sec:radiation_transport});
    the red shaded region, bounded by dashed red vertical lines, highlights mean opacities for the approximate range of equilibrium disk temperatures $\Teq$ seen in the right half of Figure~\ref{fig:thermal_equilibrium};
    the dotted purple vertical line is at the stellar temperature $T_*$ (Section~\ref{sec:stellar-irradiated_disk}).
    Referring to the left vertical axis in the right panel, solid black and gray curves show the monochromatic absorption $\knuabs$ and scattering opacities $\knusca$, respectively (Section~\ref{sec:dust_opacities});
    bar colors correspond to the same colored mean opacities in the left panel, bar widths span the frequency ranges of $f$, and bar heights show the band-mean opacities computed at $T_*$ (i.e., the intersections of $T_*$ and each $\kappa$ in the left panel).
    Referring to the right vertical axis in the right panel, the red shaded region shows Planck spectra $B_\nu$ [equation~\eqref{eq:B_nu}] for the range of $\Teq$ in the left panel and the dotted purple curve that for $T_*$.}
    \label{fig:opacities}
\end{figure*}

Our Python script (Appendix~\ref{appx:public_repository}) precomputes the opacity coefficients for each frequency band $f$ at the temperature points $T \in [\Tmin, \Tmax]$ specified in the Athena++ input file.\footnote{The input parameter sets the number of temperature points, which are distributed logarithmically between $\Tmin$ and $\Tmax$.}
For the Planck mean absorption $\kpf$ and $\kpfe$,\footnote{
These respective coefficients of $\varepsilon_{0,f}$ and $J_{0,f}$ are grouped as the last two terms on the right-hand side of equation~\eqref{eq:comoving_sources}.
Section~\ref{sec:irradiative--thermal_equilibrium} explains how their evaluation can differ by temperature.}
Rosseland mean absorption $\kaf$, and pure-scattering mean $\ksf$ as \citep[cf.][eq.~14]{Jiang2022}
\begin{align}
    \label{eq:kpfe}
    \kappa_{\mathrm{p},f(,\mathrm{e})}(T) &\equiv \frac{\int_{\nu_f}^{\nu_{f+1}}\knuabs B_\nu(T)\od\nu}{\int_{\nu_f}^{\nu_{f+1}}B_\nu(T)\od\nu}, \\
    \label{eq:krf}
    \frac{1}{\krf(T)} &\equiv \frac{\int_{\nu_f}^{\nu_{f+1}}(\knuabs + \knusca)^{-1}(\partial B_\nu/\partial T)\od\nu}{\int_{\nu_f}^{\nu_{f+1}}(\partial B_\nu/\partial T)\od\nu}, \\
    \label{eq:ksf}
    \frac{1}{\ksf(T)} &\equiv \frac{\int_{\nu_f}^{\nu_{f+1}}(\knusca)^{-1}(\partial B_\nu/\partial T)\od\nu}{\int_{\nu_f}^{\nu_{f+1}}(\partial B_\nu/\partial T)\od\nu}, \\
    \label{eq:kaf}
    \kaf(T) &\equiv \krf(T) - \ksf(T),
\end{align}
where monochromatic absorption $\knuabs$ and scattering $\knusca$ opacities must be provided as tabular input.
The script uses the Rayleigh--Jeans law for $h\nu/\kB T \ll 1$ \citep[][eq.~1.53]{RybickiLightman1979} and Wien law for $h\nu/\kB T \gg 1$ \citep[][eq.~1.54]{RybickiLightman1979} to help avoid floating-point underflow when evaluating equation~\eqref{eq:B_nu} or $\partial B_\nu/\partial T$ \citep[][eq.~1.55]{RybickiLightman1979}.
It also accounts for invalid values that may still result when evaluating equations~\eqref{eq:kpfe} through \eqref{eq:krf} due to limitations in numerical precision at the asymptotic limits of $h\nu/\kB T$.
For the particular $\knuabs$ and $\knusca$ detailed in Section~\ref{sec:dust_opacities}, examples of the tabulated mean opacities outputted by the script are shown in Figure~\ref{fig:opacities} for $N_f = 4$.
Finally, our Athena++ problem generator (Appendix~\ref{appx:public_repository}) can read these outputted tables and can linearly interpolate them as a function of the local thermal temperature $\Ttherm$ (cf. Section~\ref{sec:irradiative--thermal_equilibrium}) for each cell and for each frequency band.

\subsection{Irradiative--Thermal Equilibrium} \label{sec:irradiative--thermal_equilibrium}

The temperature dependence of Planck's law requires careful consideration when used as a weighting function for mean opacities under a multigroup framework.
For a wide band $f$ (e.g., where $\nu_{f+1}/\nu_f \gtrapprox 3$), the mean opacities defined by equations~\eqref{eq:kpfe} through \eqref{eq:ksf} can vary by an order of magnitude or more as the frequency at which $\max[B_\nu(T)]$ occurs \citep[cf.][eq.~1.56a]{RybickiLightman1979},
\begin{equation}
    \nupeak \approx \frac{2.82}{h}\kB T,
    \label{eq:nupeak}
\end{equation}
displaces across $f$ as a linear function of temperature.
This can be seen for $\kpf(T)$, $\krf(T)$, and $\ksf(T)$ in the left panel of Figure~\ref{fig:opacities} for bands $1 \leq f \leq 3$.
Given this potentially immense range in mean opacities for a wide band (e.g., when using a limited number of frequency bands $N_f$), we check for \textit{irradiative--thermal equilibrium} to help ensure those used in optically thin regions (e.g., the rarefied yet irradiated disk atmosphere, where the local thermal temperature may be hundreds of Kelvin) are not under weighted by Planck's law (e.g., evaluated significantly below the effective stellar temperature at thousands of Kelvin).

Irradiative--thermal equilibrium is met when the frequency at the spectral peak of the radiation field is close to $\nupeak$ for the local thermal temperature $\Ttherm$.
In other words, $\Ttherm$ should be similar to the \textit{color temperature} $\Tc$ which describes an indeterminate blackbody source of unknown absolute scale or distance \citep[e.g., the host star;][p.~26]{RybickiLightman1979}.
Note that local thermodynamic equilibrium neither guarantees irradiative--thermal equilibrium ($\Tc \approx \Ttherm$) nor is guaranteed by it \citep[][p.~32]{RybickiLightman1979}.

To assess irradiative--thermal equilibrium in a given cell, our Athena++ problem generator (Appendix~\ref{appx:public_repository}) compares the frequency bands containing the spectral peaks of the specific radiation energy density $\Erf/\Delta\nu_f$ and a localized blackbody at $\Ttherm$ (i.e., $\firrad \stackrel{?}{=} \ftherm$).
A subroutine determines $\firrad$ as the band $0 \leq f < N_f - 1$ that contains $\max(\Erf/\Delta\nu_f)$, where the radiation energy density \citep[cf.][eq.~7]{Jiang2022}
\begin{equation}
    \Erf = 4\pi\sum_{n=0}^{\Nang-1}I_f(\nhat)w_n
    \label{eq:Erf}
\end{equation}
for the specific intensity along each angle $I_f(\nhat)$, and where the bandwidth $\Delta\nu_f = \nu_{f+1} - \nu_f$ (Section~\ref{sec:radiation_transport}).
The highest frequency band $f = N_f - 1 = [\nu_{N_f - 1}, \infty)$ is excluded from the search, as $E_{\mathrm{r},N_f - 1}/\Delta\nu_{N_f - 1}$ always evaluates to zero.
To identify $\ftherm$, the Wien displacement law [equation~\eqref{eq:nupeak}] is used to compute $\nupeak(\Ttherm)$ before another subroutine identifies finds $f$ where $\nu_f < \nupeak(\Ttherm) < \nu_{f+1}$.

If $\firrad = \ftherm$, which typically occurs in optically thick regions [e.g., the dense disk midplane shielded from stellar irradiation; Section~\ref{sec:stellar-irradiated_disk} and equation~\eqref{eq:tau_*}], the code interpolates the absorption opacity coefficient $\kpfe$ [of the second mean radiation energy density $J_{0,f}$ on the right-hand side of equation~\eqref{eq:comoving_sources}] at $\Ttherm$, as usual (Section~\ref{sec:emissivity_opacity}).
However, if $\firrad \neq \ftherm$, the code interpolates $\kpfe$ at $\Tc$ instead (e.g., closer to the effective stellar temperature $T_*$).
We estimate $\Tc$ by solving equation~\eqref{eq:nupeak} for $T$ but use the arithmetic center of $\firrad$ as $\nupeak = (\nu_{\firrad} + \nu_{\firrad+1})/2$.
As the error in estimating $\Tc$ scales with the bandwidth $\Delta\nu_{\firrad}$, the overall condition becomes less important for narrow bands or higher $N_f$ since
\begin{equation}
    \lim_{N_f\to\infty}
    \begin{cases}
        \kappa_{\mathrm{p},f(,\mathrm{e})} = \kaf = \knuabs,\\
        \ksf = \knusca
        \label{eq:kf-to-knu_limit}
    \end{cases}
\end{equation}
[see also Section~\ref{sec:dust_opacities} and equation~\eqref{eq:f_*-to-nu_*_limit}].

\section{Model} \label{sec:model}

This section establishes the stellar irradiated disk model and dust opacities we use to demonstrate the methodology in Section~\ref{sec:methodology} and to compare with other radiative transfer methods and codes in Section~\ref{sec:hydrostatic_comparisons}.
Section~\ref{sec:stellar-irradiated_disk} formulates the axisymmetric disk profiles and specifies the key parameters for stellar irradiation.
Section~\ref{sec:dust_opacities} details the monochromatic and mean dust opacities.
Finally, Section~\ref{sec:thermal_equilibrium} differentiates how our hydrostatic models reach thermal equilibrium.

\subsection{Stellar-Irradiated Disk} \label{sec:stellar-irradiated_disk}

The computational domain is in spherical coordinates $(r, \theta, \phi)$.
As specified in our Athena++ input file (Appendix~\ref{appx:public_repository}), it extends radially from $\rmin = 10\au$ to $\rmax = 100\au$, polarly from $\thetamin = 0$ to $\thetamax = \pi$ (corresponding to meridional latitudes from $90\degree$ to $-90\degree$ with the disk midplane at $0\degree$), and azimuthally from $\phimin = 0$ to $\phimax = 2\pi$.
Except for a brief radial-resolution study in Section~\ref{sec:without_dust_emission}, we use $N_r = 128$ logarithmically spaced grid cells radially, $N_\theta = 1024$ uniformly spaced cells polarly,\footnote{
Given the range in $\theta$ for our domain, we find $N_\theta = 1024$ better distinguishes radial profiles (Section~\ref{sec:hydrostatic_comparisons}) at intermediate latitudes (e.g., $10\degree$) after trying as few as 128 cells.}
and $N_\phi = 1$ with periodic boundary conditions azimuthally for axisymmetry.

The disk profiles in our problem generator (Appendix~\ref{appx:public_repository}) are adapted from those developed by \cite{ZhuDongStone2015} for their \texttt{disk.cpp} problem generator included in the public version of Athena++.
Converting from cylindrical coordinates $(R, \phi, Z)$, where $r = \sqrt{R^2 + Z^2}$ and cylindrical radius $R = r\sin\theta$, the disk density from \citet[][cf. eq.~5]{FromangLyraMasset2011} becomes
\begin{equation}
    \rho(r, \theta) = \rhosig\exp\left(\frac{GM_*}{\csiso^2r}\left[1 - \frac{1}{\sin\theta}\right]\right),
    \label{eq:rho}
\end{equation}
where $G$ is the gravitational constant, $M_*$ is the central stellar mass, and
\begin{equation}
    \csiso^2(r, \theta) = \csison^2\left(\frac{r\sin\theta}{R_0}\right)^p,
    \label{eq:csiso}
\end{equation}
is the locally isothermal sound speed squared as a power law in $R$ \citep[cf.][eq.~1]{FromangLyraMasset2011} with index $p = -1/2$.
In the thin-disk limit ($\theta \simeq \thetamid \equiv \pi/2$), expanding equation~\eqref{eq:rho} to first order yields \citep[cf.][eq.~3, where $Z = r\cos\theta$]{TakeuchiLin2002}
\begin{equation}
    \rho(r, \theta \simeq \thetamid) \simeq \rhosig\exp\left(-\frac{[r\cos\theta]^2}{2H^2}\right),
    \label{eq:rho_lim}
\end{equation}
where the disk scale height \citep[cf.][eq.~4]{TakeuchiLin2002}
\begin{equation}
    H(r, \thetamid) \equiv \frac{\csiso}{v_\phi}r,
    \label{eq:H}
\end{equation}
for Keplerian velocity $v_\phi(r) = \sqrt{GM_*/r}$.
Setting $\csison^2/v_\phi^2(R_0) = 4.80\times10^{-3}$, the disk aspect ratio at $R_0 = 17.0\au$ is $H_0/R_0 \approx 0.07$.

The extinction of stellar irradiation (Section~\ref{sec:stellar_irradiation}) near the inner radial boundary should be carefully considered when using the implicit solver \citep[][\S~3]{Jiang2021} for the radiation transport module (Section~\ref{sec:radiation_transport}).
The \textit{effective} optical depth to peak stellar irradiation \citep[cf.][eqs.~1.26 and 1.98]{RybickiLightman1979}
\begin{equation}
    \tau_*(r, \theta, \phi) = \int_{\rmin}^{r}\rho(r', \theta, \phi)(\kstarabs + \kstarsca)\od r',
    \label{eq:tau_*}
\end{equation}
where $\rho(\kstarabs + \kstarsca)$ is the extinction coefficient \citep[cf.][p.~36 and eq.~1.22]{RybickiLightman1979} integrated along a radial path $\od r'$ at fixed $\theta$ and $\phi$ in the computational domain.
Evaluating equation~\eqref{eq:nupeak} at the effective stellar temperature $T_*$ yields $\nu_*$, the frequency at which the monochromatic absorption and scattering opacity coefficients (Section~\ref{sec:dust_opacities}) correspond to $\kstarabs$ and $\kstarsca$, respectively.
A medium is effectively thin or translucent
if $\tau_* \ll 1$ and effectively thick or opaque if $\tau_* \gg 1$ \citep[][p.~38]{RybickiLightman1979}.
An inner radial boundary opaque to stellar irradiation can cause the implicit solver to require an excessive number of iterations---sometimes reaching the iteration limit---during many of the initial time steps.
The difficulty arises because the radiation field can change substantially from one time step to the next near such a boundary, thereby increasing the number of iterations required by the implicit solver \citep[Section~\ref{sec:performance} and][\S~3.2.2]{Jiang2021}.
As discussed in Section~\ref{sec:applications}, one way to mitigate this behavior is to ramp the opacity coefficients up to their correct values over a fixed radial softening length, thereby reducing the optical depth near the radiative boundary.
However, because this approach cannot be implemented in the Monte Carlo radiative transfer code (Section~\ref{sec:radmc-3d}) used for comparison in Section~\ref{sec:hydrostatic_comparisons}, we do not adopt it here.

\begin{figure}
    \includegraphics[width=\columnwidth]{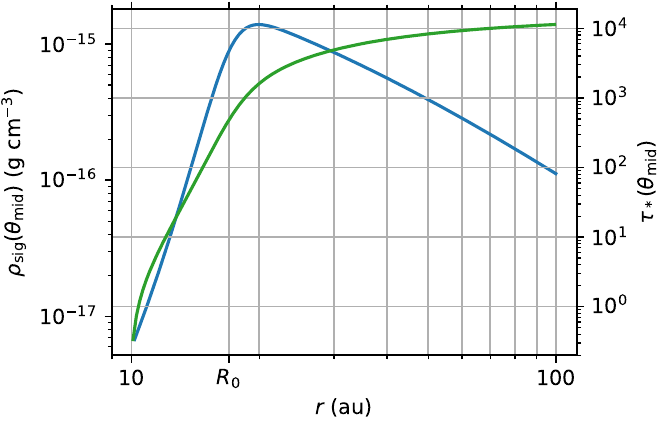}
    \caption{Midplane radial profiles at $\thetamid \equiv \pi/2 = 0\degree$ latitude (cf. Figure~\ref{fig:vertical}).
    Referring to the left vertical axis, the blue curve shows the density $\rhosig$ [equation~\eqref{eq:rhosig}].
    Referring to the right axis, the green curve shows the effective depth to peak stellar irradiation $\tau_*$ [equation~\eqref{eq:tau_*}].}
    \label{fig:midplane}
\end{figure}

An alternate way to make the inner boundary effectively translucent to stellar irradiation is to soften the radial density profile which also softens the extinction coefficient.
Thus, we multiply the standard density power law in cylindrical $R$ \citep[][eq.~3]{NelsonGresselUmurhan2013} by a logistic (i.e., sigmoid) function to express 
\begin{equation}
    \rhosig(r, \theta) = \frac{\rho_0[(r\sin\theta + R_0)/R_0]^d}{1 + \exp\left(-e^e[(r\sin\theta - R_0)/R_0]\right)},
    \label{eq:rhosig}
\end{equation}
where $R_0 = 17.0\au$ and we set $d = -9/4$. 
As a result, the midplane radial profile $\rhosig(r, \thetamid)$ shown in Figure~\ref{fig:midplane} increases by more than two orders of magnitude across $10\au$, peaking smoothly around $r \approx 20\au$.
For the opacities detailed in Section~\ref{sec:dust_opacities}, the corresponding extinction coefficient provides an effectively thin inner boundary, with $\tau_*(r \approx \rmin, \thetamid) < 1$ also shown in Figure~\ref{fig:midplane}.
This density profile improves the performance of the implicit solver \citep[][\S~3.2.2]{Jiang2021} and can be implemented in a Monte Carlo radiative transfer code (Section~\ref{sec:radmc-3d}) to compare hydrostatic models (Section~\ref{sec:hydrostatic_comparisons}). We defer a discussion of its physical implications (e.g., as a transition disk) to Section~\ref{sec:applications}.

\begin{figure}
    \includegraphics[width=\columnwidth]{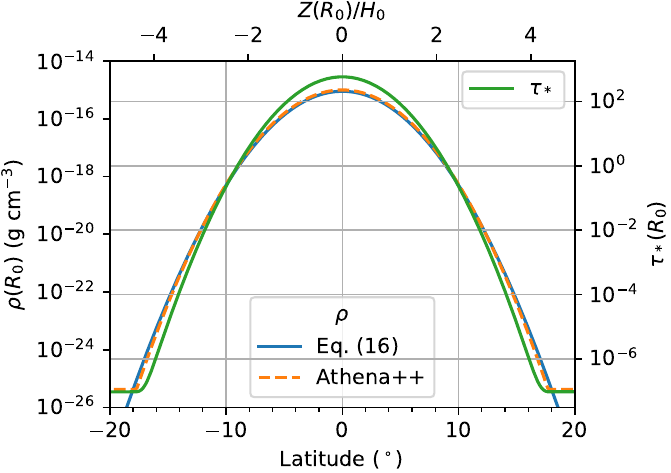}
    \caption{Meridional profiles at $r = R_0 = 17.0\au$  (cf. Figure~\ref{fig:midplane}).
    The upper horizontal axis converts latitudes along the bottom to $Z(R_0) = R_0\cos\theta$ in scale heights $H_0$ at $R_0$ (Section~\ref{sec:stellar-irradiated_disk}).
    Referring to the left vertical axis, the solid blue curve shows the density $\rho$ in the thin-disk limit [equation~\eqref{eq:rho_lim}] and the dashed orange the Athena++ output [equation~\eqref{eq:rho}] with floor $\rhomin = 4.28\times10^{-26}\gm\cm^{-3}$.
    Referring to the right axis, the solid green curve shows the effective optical depth to stellar irradiation $\tau_*$ [equation~\eqref{eq:tau_*}] as outputted by the code.}
    \label{fig:vertical}
\end{figure}

\begin{figure*}
    \includegraphics[width=\textwidth]{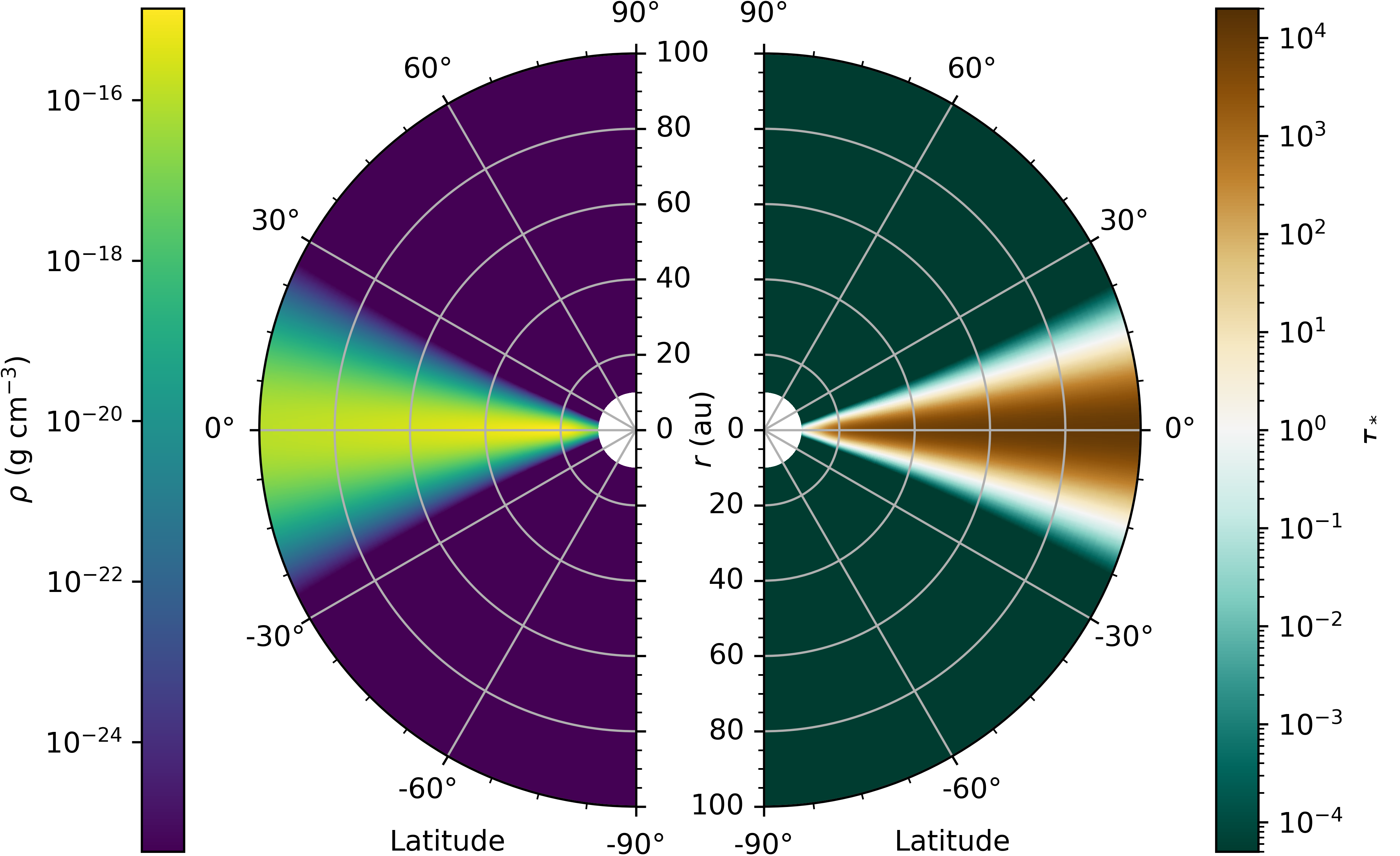}
    \caption{The axisymmetric protoplanetary disk model.
    In this meridional projection, radius $r$ is in astronomical units (au), and disk latitudes are in degrees with the midplane at $0\degree$ and the poles at $\pm90\degree$.
    The left half shows the time-independent (Sections~\ref{sec:thermal_equilibrium}) dust density $\rho$ [equation~\eqref{eq:rho}] and the right the effective optical depth to peak stellar irradiation $\tau_*$ as integrated radially outward [equation~\eqref{eq:tau_*}].}
    \label{fig:disk}
\end{figure*}

Setting $\rho_0 = 8.56\times10^{-15}\gm\cm^{-3}$ in equation~\eqref{eq:rhosig} sets the absolute scales of the left vertical axes in Figures~\ref{fig:midplane} and \ref{fig:vertical}.
The latter shows meridional density profiles at $r = R_0$ as approximated by equation~\eqref{eq:rho_lim} and as outputted by Athena++ based on equation~\eqref{eq:rho}.
Figure~\ref{fig:vertical} also plots the effective depth to peak stellar irradiation as outputted by the code and shows $\tau_* \approx 1$ at $\pm9\degree$ in latitude (just above $2H_0$ in the upper horizontal axis).
The abrupt flattening of the outputted profiles above and below $\pm18\degree$, respectively (just above $4H_0$), results from enforcing a density floor $\rhomin = 4.28\times10^{-26}\gm\cm^{-3}$.
These density parameters also set the absolute scale of the color bar for the axisymmetric meridional map of equation~\eqref{eq:rho} in the left half of Figure~\ref{fig:disk} where most of the space is filled at this floor density.

\begin{figure*}
    \includegraphics[width=\textwidth]{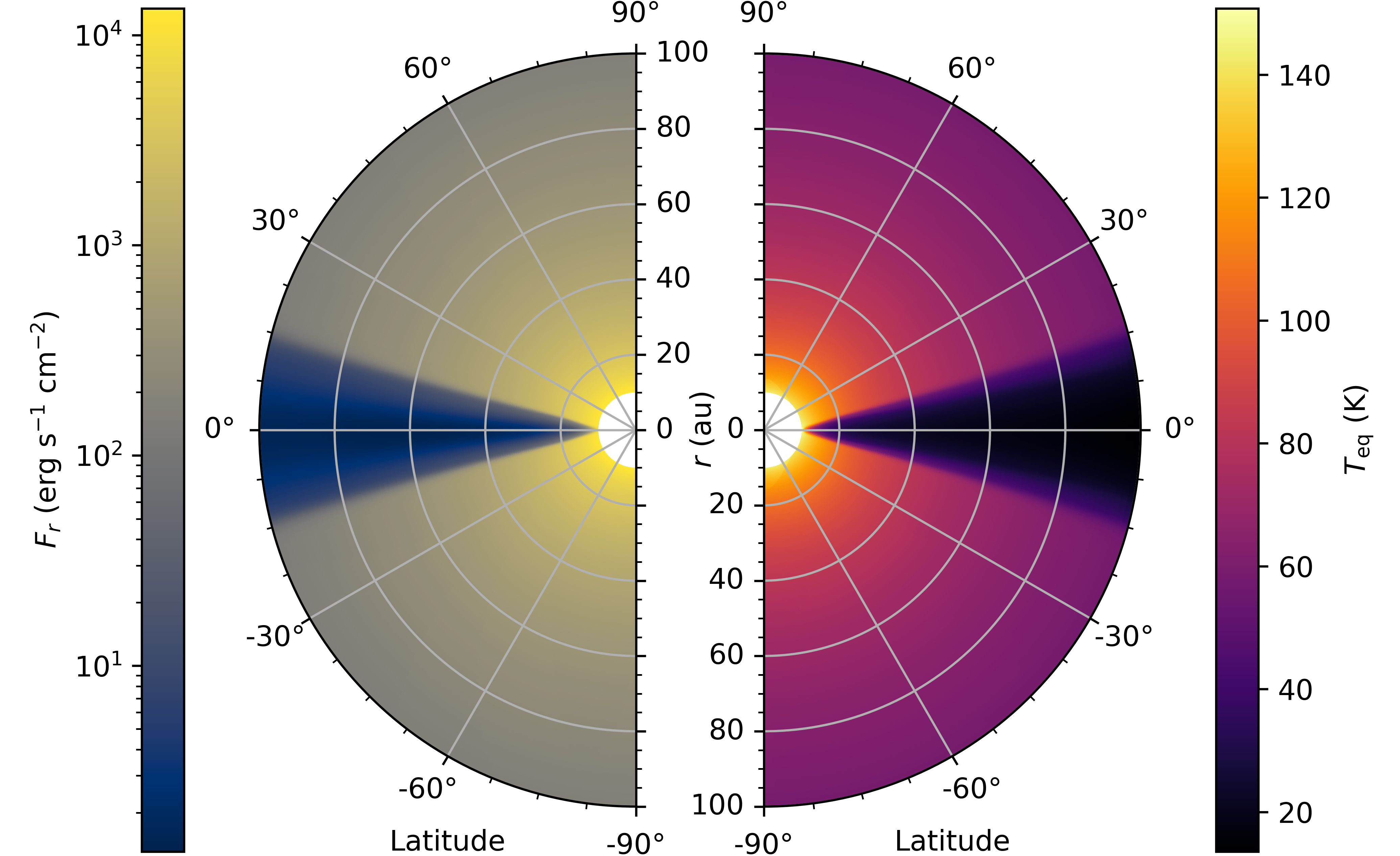}
    \caption{Similar to Figure~\ref{fig:disk} except at thermal equilibrium (cf. Figure~\ref{fig:evolution}) for the hydrostatic model that includes frequency-dependent absorption and scattering opacities across $N_f = 64$ bands.
    The left half shows the frequency-integrated radial flux $F_r$ [equation~\eqref{eq:Fr}] and the right the equilibrium temperature $\Teq$ (Section~\ref{sec:thermal_equilibrium}).}
    \label{fig:thermal_equilibrium}
\end{figure*}

At the origin of the global spherical coordinate system, we assume a sunlike star for the stellar irradiation parameters introduced in Section~\ref{sec:stellar_irradiation}
and in our Athena++ input file \citep[cf. Appendix~\ref{appx:public_repository} and][Tab.~1]{PrsaHarmanecTorres2016}.
Evaluating equation~\eqref{eq:I-rout} at the inner radial boundary $\rmin = 10\au$ yields a frequency-integrated radial flux $F_r(\rmin) = I(\rmin, \nhatrout)$ [cf. equation~\eqref{eq:Fr}] that is 1\% of the solar constant (i.e., irradiance at $r = 1\au$) due to the inverse-square law.
This approximate value can be seen as $\max(F_r)$ along $\rmin = 10\au$ in the left half of Figure~\ref{fig:thermal_equilibrium}.
Finally, we note the purple dotted lines in Figure~\ref{fig:opacities} shows $T_*$ in the left panel and $B_\nu(T_*)$ from equation~\eqref{eq:B_nu} in the right panel with $\nu_* \approx 3.40\times10^{14}\Hz$ from equation~\eqref{eq:nupeak}.

\subsection{Dust Opacities} \label{sec:dust_opacities}

To demonstrate the framework presented in Section~\ref{sec:methodology} and compare with other radiative transfer methods and codes in Section~\ref{sec:hydrostatic_comparisons}, we use the monochromatic dust opacities generated by \cite{ZhangZhuJiang2024} from the methods and Python package developed by \cite{BirnstielDullemondZhu2018} to perform Mie opacity calculations.
As used by the Disk Substructures at High Angular Resolution Project \citep[DSHARP,][]{AndrewsHuangPerez2018}, the protoplanetary disk grain structure and composition specified by \citet[][\S~2]{BirnstielDullemondZhu2018} follow \cite{PollackHollenbachBeckwith1994} but assume nonporous, spherical grains with a mass fraction of 20\% in water ice, 33\% in astronomical silicates, 7\% in troilite, and 40\% in refractory organic material.
With this DSHARP composition, \citet[][\S~2 and Tab.~1]{BirnstielDullemondZhu2018} use refractive index data from \cite{WarrenBrandt2008} for water, \cite{Draine2003} for silicates, and \cite{HenningStognienko1996} for troilite and refractory organics to calculate the size-dependent mass absorption and scattering coefficients $\knuabs(a)$ and $\knusca(a)$, respectively, for a grain with radius $a$.

For grain-size-averaged opacities, \citet[][\S~2.2.1]{ZhangZhuJiang2024} assumed a grain-size number distribution \citep[cf.][eq.~5]{BirnstielDullemondZhu2018}
\begin{equation}
    n(a) \propto
    \begin{cases}
        a^{-q}\quad&\mathrm{for}\ \amin \leq a \leq \amax,\\
        0          &\mathrm{otherwise},
    \end{cases}
    \label{eq:grain-size_distribution}
\end{equation}
with power-law index $q = 3.5$ \citep{MathisRumplNordsieck1977} and minimum and maximum grain radii $\amin = 0.1\micron$ and $\amax = 1\micron$, respectively.
The total dust density is thus $\rhodtot = \int_0^\infty n(a)m(a)\od a$, where $m(a)$ is the size-dependent grain mass.
For the parameter values we set in Section~\ref{sec:stellar-irradiated_disk}, integrating the density in equation~\eqref{eq:rho} over the volume of the computational domain yields a total dust disk mass $M \approx 1.30\times10^{-4} M_\odot$.
Thus, we can assume the density field only represents dust corresponding to this size distribution.
The weighting of $\knuabs(a)$ and $\knusca(a)$ by $\rhodtot$ \citep[][eq.~6]{BirnstielDullemondZhu2018} yields the size-averaged monochromatic opacities $\knuabs$ and $\knusca$ shown as solid black and gray lines, respectively, in the right panel of Figure~\ref{fig:opacities}.
According to \citet[][Appx.~B]{BirnstielDullemondZhu2018}, the optical properties of water account for the sharp bump in $\knuabs$ at $10^{14}\Hz$ (cf. the features at $3\micron$ in their Fig.~11), which can slightly change depending on the mass fraction assumed in the grain composition.
Ultimately, since the framework can utilize any $\knuabs$ and $\knusca$ provided, we remain agnostic as to whether the DSHARP composition is the best choice for protoplanetary disks \citep{ZagariaFacchiniCurone2025}.

Figure~\ref{fig:opacities} also plots the corresponding band-mean opacities (Section~\ref{sec:emissivity_opacity}) for $N_f = 4$, as functions of temperature\footnote{
In the Athena++ input file (Appendix~\ref{appx:public_repository}), we specify 1,000 logarithmically uniform temperature points between $\Tmin = 10\K$ and $\Tmax = 10^{4}\K$.}
and frequency in its left and right panels, respectively.
The right panel is useful in identifying $f_*$, the band containing the peak stellar frequency $\nu_*$ at which $\max[B_\nu(T_*)]$ occurs.
This identification matters because equation~\eqref{eq:tau_*} defines the effective optical depth $\tau_*$ using the monochromatic extinction at $\nu_*$, whereas the multigroup calculation only has access to band-mean opacities.
Thus, the mean opacities in band $f_*$ at $T_*$ is a multigroup approximation to $\kstarabs$ and $\kstarsca$ (see also Section~\ref{sec:irradiative--thermal_equilibrium}).
As the number of frequency bands increases, the width of $f_*$ decreases, such that the approximation converges toward the monochromatic value used in equation~\eqref{eq:tau_*}, i.e., equation~\eqref{eq:kf-to-knu_limit} and
\begin{equation}
    \lim_{N_f\to\infty} f_* = \nu_*.
    \label{eq:f_*-to-nu_*_limit}
\end{equation}
We explore the relationship between $f_*$ and $\tau_*$ in the frequency-resolution study of Section~\ref{sec:frequency_bands}.

For the $N_f = 64$ model, we show an axisymmetric meridional map of the approximate $\tau_*$ in equation~\eqref{eq:tau_*}  in the right half of Figure~\ref{fig:disk} and highlight three distinct regions: (1) the effectively thin disk atmosphere, above and below latitudes of $\pm20\degree$, respectively; (2) the effectively thick midplane region within $\pm10\degree$; and (3) an intermediate region between these latitudes, where $\tau_* \sim 1$.

\subsection{Thermal Equilibrium} \label{sec:thermal_equilibrium}

To compare with irradiated disk models from Monte Carlo radiative transfer calculations (Section~\ref{sec:hydrostatic_comparisons}), we modified the Athena++ source code (Appendix~\ref{appx:public_repository}) to not add flux divergence to the hydrodynamic variables during the integration time-step routine.
For consistency and to improve the performance of the iterative implicit solver \citep[][\S~3.2.2]{Jiang2021}, we also remove radiation pressure as a source of momentum for the fluid.
Thus, the density equation~\eqref{eq:rho} and its field, shown in the left half of Figure~\ref{fig:disk}, become time independent.
To further improve the iterative convergence of the implicit solver (as explained in more detail at the end of Section~\ref{sec:optimization}), we also reduce the speed of light to 1\% in all our models, which does not change the steady-state solution.

\begin{figure}
    \includegraphics[width=\columnwidth]{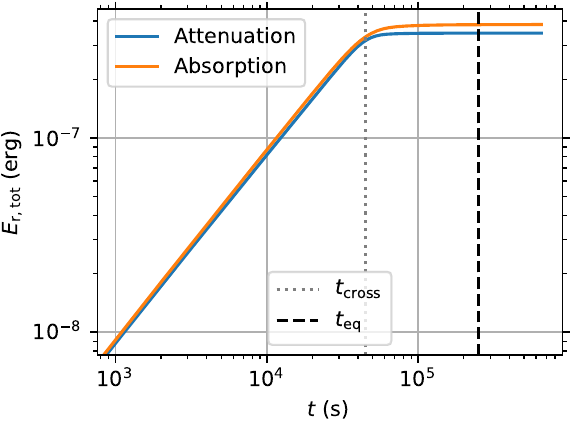}
    \caption{Total radiation energy density $\Ertot$ [equation~\eqref{eq:Ertot}] as a function of time for hydrostatic models that include frequency-dependent absorption and scattering opacities across $N_f = 64$ bands.
    The blue curve shows the case for radiative attenuation only (i.e., without dust heating), the orange shows that for absorption (i.e., with dust thermal reemission), the dotted gray vertical line shows the light-crossing time $\tcross$ for the 90~au radial domain (Section~\ref{sec:thermal_equilibrium}), and the dashed black vertical line marks when thermal equilibrium is reached at $\teq$ (Section~\ref{sec:performance}).
    Note that $t$ is multiplied by 1\% (our \texttt{reduced\_factor}) to account for using a reduced speed of light (Sections~\ref{sec:thermal_equilibrium} and \ref{sec:optimization}).}
    \label{fig:evolution}
\end{figure}

One way to assess the evolution of the radiation field is to measure the radiative energy.
Summing equation~\eqref{eq:Erf} over all frequency bands gives the integrated radiative energy density \citep[cf.][eq.~8]{Jiang2022}
\begin{equation}
    \Er(r, \theta, \phi) = \sum_{f=0}^{N_f}\Erf(r, \theta, \phi),
    \label{eq:Er}
\end{equation}
and integrating this over the entire domain gives the total radiative energy
\begin{equation}
    \Ertot = \iiint\Er\od r\od\theta\od\phi.
    \label{eq:Ertot}
\end{equation}
Figure~\ref{fig:evolution} plots the evolution of $\Ertot$ for two hydrostatic models.
As stellar radiation from the inner boundary (Section~\ref{sec:stellar_irradiation}) propagates radially throughout the computational domain (Section~\ref{sec:stellar-irradiated_disk}), $\Ertot$ increases almost linearly with time in both cases before turning over at the light-crossing time $\tcross$ for $90\au$.

Whether stellar irradiation is allowed to heat the dust accounts for the slight difference in $\Ertot(t > \tcross)$ between the two models in Figure~\ref{fig:evolution}.
If the \texttt{<radiation>} parameter \texttt{source\_flag = 0} in the Athena++ input file (Appendix~\ref{appx:public_repository}), stellar irradiation is attenuated by any opacity but does not heat the matter and thus remains the only source of radiative energy such that $\Ertot$ reaches a maximum at $\tcross$.
If \texttt{source\_flag > 0},\footnote{
We use \texttt{source\_flag = 2}, but \texttt{source\_flag = 1} additionally ensures the internal energy will not become negative.}
the irradiated dust also heats up according to its absorption opacity and reemits as a blackbody (Section~\ref{sec:emissivity_opacity}), whose thermalization of which accounts for the continued increase in $\Ertot(t > \tcross)$ toward a higher maximum at $\teq$ (Section~\ref{sec:performance}) but over a longer diffusion timescale.

Another way to inspect the radiation field is to measure the radiative flux.
Summing the net radial flux (cf. eq.~7 of \citealt{Jiang2022} and eq.~1.3b of \citealt{RybickiLightman1979})
\begin{equation}
    \Frf(r, \theta, \phi) = 4\pi\sum_{n=0}^{\Nang-1}I_f(r, \theta, \phi, \nhat)\cdot\hat{\mathbf{r}}w_n,
    \label{eq:Frf}
\end{equation}
over all frequency bands gives the integrated net radial flux (cf eq.~8 of \citealt{Jiang2022} and eq.~1.5a of \citealt{RybickiLightman1979})
\begin{equation}
    F_r(r, \theta, \phi) = \sum_{f=0}^{N_f}\Frf(r, \theta, \phi).
    \label{eq:Fr}
\end{equation}
The left half of Figure~\ref{fig:thermal_equilibrium} shows an axisymmetric meridional map of $F_r$ for $N_f = 64$ at thermal equilibrium ($t > \teq$).
Referring to the radial optical depth field in the right half of Figure~\ref{fig:disk}, $F_r \propto r^{-2}$ in the effectively thin atmosphere (e.g., meridional latitudes above and below $\pm15\degree$, respectively) where $\tau_* < 1$ [equation~\eqref{eq:tau_*}].
Meanwhile, $F_r$ is attenuated by more than an order of magnitude toward the midplane (e.g., within $\pm15\degree$) where $\tau_* > 1$.

Comparing each half of Figure~\ref{fig:thermal_equilibrium} shows the direct correlation between the irradiative flux on the left and the equilibrium dust temperature $\Teq$ on the right.
The effectively thick region near the midplane keeps the dust very cold, on the order of $10\K$.
Although the atmosphere is tenuous and effectively thin overall, dust opacities skewed toward higher frequencies allow sparse grains to absorb ultraviolet stellar irradiation [cf. $\knuabs$ and $B_\nu(T_*)$ in Figure~\ref{fig:opacities}] and heat up to above $100\K$.

\section{Hydrostatic Comparisons} \label{sec:hydrostatic_comparisons}

To assess the accuracy of our Athena++ framework for stellar-irradiated disks, this section compares the thermal equilibrium reached in each of the hydrostatic models (Section~\ref{sec:thermal_equilibrium}) with those produced by other radiative transfer methods.
Section~\ref{sec:comparative_methods} introduces these other methods.
Sections~\ref{sec:gray_opacity} and \ref{sec:frequency-dependent_opacities} compare frequency-independent and frequency-dependent models, respectively.
Section~\ref{sec:resolution_studies} studies the convergence and computational performance of our framework as a function of the resolution parameters for the radiation transport module.
These studies provide context and motivation for an optimized model demonstrated in Section~\ref{sec:optimization}.

\subsection{Comparative Methods} \label{sec:comparative_methods}

We first introduce the other radiative transfer methods we use to produce the models with which we compare our framework.
Section~\ref{sec:radmc-3d} introduces a Monte Carlo radiative transfer code.
Section~\ref{sec:hybrid_ray_tracing} introduces an alternative hybrid ray-tracing method for Athena++.

\subsubsection{RADMC-3D} \label{sec:radmc-3d}

To compare the equilibrium dust temperatures of stellar irradiated disk models (Section~\ref{sec:thermal_equilibrium}), we use RADMC-3D \citep{DullemondJuhaszPohl2012} which implements the Monte Carlo radiative transfer method from \cite{BjorkmanWood2001} improved with the continuous absorption method from \cite{Lucy1999}.
We use the discretized axisymmetric domain and fixed density field [equation~\eqref{eq:rho}] outputted by Athena++ as inputs for RADMC-3D to ensure there are no differences in machine precision between spherical grids or disk setups.
Similarly, we treat the star as a point source centered at the origin with the same stellar properties, e.g., $R_*$ and $T_*$ (Section~\ref{sec:stellar-irradiated_disk}).

We use $8\times10^8$ photon packets to reduce the inherent noise of Monte Carlo solutions.
The wavelengths of these photons are randomly drawn from 150 equally weighted logarithmically spaced points with 20 between $0.1\micron$ and $7\micron$, 100 between $7\micron$ and $25\micron$, and 30 between $25\micron$ and $10^4\micron$.
As the trajectories of these photons are followed within the domain, isotropic thermal reemission (immediately when absorbed) and isotropic scattering angles for each interaction are randomly drawn from a uniform distribution.\footnote{
For the simplified anisotropic scattering discussed in Section~\ref{sec:scattering}, the angular distribution of scattered photons is weighted according to the $g$ parameter \citep{HenyeyGreenstein1941}.}
We use the same monochromatic DSHARP opacities $\knuabs$ and $\knusca$ detailed in Section~\ref{sec:dust_opacities} (and plotted in Figure~\ref{fig:opacities}) for the comparisons in Section~\ref{sec:frequency-dependent_opacities}.
For each RADMC-3D model, we developed a Python script (Appendix~\ref{appx:public_repository}) to create and format the required input files accordingly.

\subsubsection{Hybrid Ray Tracing} \label{sec:hybrid_ray_tracing}

In Sections~\ref{sec:gray_with_dust_emission} and \ref{sec:with_dust_emission}, we also compare to hybrid ray tracing in Athena++.
Appendix~\ref{appx:hybrid_ray-tracing_method} details this method for frequency-dependent stellar irradiation, following a similar approach by \cite{KuiperKlahrDullemond2010}.
Unlike our framework, it does not support radiation pressure from the star \citep[cf.][\S~3.3.4]{KuiperKlahrDullemond2010} or scattering, as the latter makes the transfer equation for ray tracing integrodifferential without a closed-form solution \citep[][\S~1.7]{RybickiLightman1979};
and dust reemission and subsequent thermalization are frequency independent with a temperature-dependent mean opacity \citep[][Fig.~1]{ZhangZhuJiang2024}.

\subsection{Gray Opacity} \label{sec:gray_opacity}

As a baseline, we first compare `gray' models where the opacity is independent of frequency.
For the Athena++ models in this section, we use $N_f = 1$ frequency band (Section~\ref{sec:radiation_transport}) and omit scattering (Section~\ref{sec:emissivity_opacity}).
Section~\ref{sec:without_dust_emission} compares only solutions without dust emission, and Section~\ref{sec:gray_with_dust_emission} compares only models with dust emission.

\subsubsection{Without Dust Emission} \label{sec:without_dust_emission}

When irradiated dust does not heat up to become an additional source of radiation (Section~\ref{sec:thermal_equilibrium}), the pure attenuation of stellar irradiation due to the absorption opacity of the dust can be solved analytically.
The formal solution to the radiative transfer equation for pure attenuation is \citep[cf.][eq.~1.29]{RybickiLightman1979}
\begin{equation}
    I(\tau) = I(0)e^{-\tau}
    \label{eq:formal_transfer_solution}
\end{equation}
where the optical depth to stellar irradiation $\tau_*$ [equation~\eqref{eq:tau_*}] reduces to $\tau$ for a gray absorption opacity $\kabs$.
Comparing equations~\eqref{eq:Frf} and \eqref{eq:Fr}, the integrated radial flux $F_r = \Frf$ when $N_f = 1$.
Substituting equation~\eqref{eq:formal_transfer_solution} into this gray $F_r$, with equation~\eqref{eq:I-rout} at the inner radial boundary as the initial intensity $I(0)$, yields the analytic ray-tracing solution for the flux of attenuated stellar irradiation [cf. equation~\eqref{eq:I-rout}]
\begin{equation}
    F_r(\tau) = \sigma T_*^4\left(\frac{R_*}{r}\right)^2e^{-\tau}.
    \label{eq:F_r-tau}
\end{equation}

\begin{figure}
    \includegraphics[width=\columnwidth]{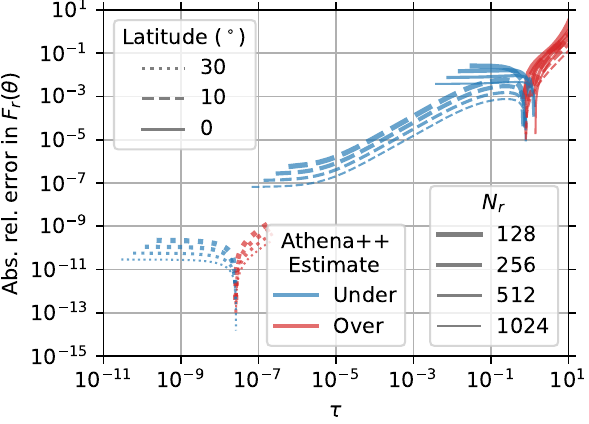}
    \caption{Absolute relative (i.e., fractional) error in radial flux $F_r$ as a function of radial optical depth $\tau$ compared to analytic ray tracing [equation~\eqref{eq:F_r-tau}] for gray models with pure radiative attenuation only (i.e., without dust emission).
    Line styles correspond to radial profiles at different disk latitudes (cf. Figure~\ref{fig:thermal_equilibrium}), weights show models with differing numbers of radial grid cells $N_r$ which are logarithmically spaced, and blue or red colors indicate where Athena++ underestimates or overestimates the analytic solution, respectively.}
    \label{fig:attenuation}
\end{figure}

Figure~\ref{fig:attenuation} shows the relative (i.e., fractional) error in $F_r(\tau)$ between the Athena++ solution and equation~\eqref{eq:F_r-tau} at latitudes representing the three distinct regions in optical depth (Section~\ref{sec:dust_opacities}; cf. Figure~\ref{fig:disk}).
For $\tau$ in this comparison, we adopt a gray absorption opacity $\kabs = 4000\cm^2\gm^{-1}$ based on the $N_f = 1$ Planck mean $\kpf(T_*)$ from equation~\eqref{eq:kpfe}.
Although $\tau$ reaches $10^3$ at the midplane ($0\degree$ latitude), we do not show the full range because Athena++ outputs $F_r(\tau \gtrsim 10) = 0$ for these models.
At each latitude, the radial flux in Athena++ underestimates and overestimates the analytic solution for relatively lower and higher values of $\tau$, respectively. 
Although the absolute relative error tends to increase with optical depth, it is also inversely proportional to the radial resolution overall (cf. $N_r = 128$ and 1024), since $\tau$ is a function of $r$ [equation~\eqref{eq:tau_*}].
These absolute trends may be related to the inverse proportionality between the mean-free path of photons and the extinction coefficient \citep[][eq.~1.93]{RybickiLightman1979};
in other words, higher grid resolutions more accurately capture radiative transfer in optically thick regions.

The findings at either end of the optical depths in Figure~\ref{fig:attenuation} warrant further discussion.
Recalling that the density of the disk atmosphere is at a constant floor $\rhomin$ (Section~\ref{sec:stellar-irradiated_disk}), the comparisons at the lowest optical depths (i.e., at $30\degree$ latitude) effectively correspond to 1D radiative transfer solutions for a uniform medium with constant density and opacity [cf. equation~\eqref{eq:tau_*}].
Meanwhile, at the highest optical depths, even if Athena++ overestimates $F_r$ by a factor of two relative to the analytic solution (i.e., by a relative error of unity) the expected midplane temperature would only increase by about 20\% ($2^{1/4} \approx 1.19$), since $T \propto F^{1/4}$ [equation~\eqref{eq:F_r-tau}].

\begin{deluxetable*}{lccclclrrrrrr}
    \tablecaption{Hydrostatic Comparisons\label{tab:hydrostatic_comparisons}}
    \tablecolumns{13}
    \tablehead{
        \colhead{Opacity$\anote$}   & \colhead{Section}                 & \colhead{Figure}          & \colhead{Latitude$\bnote$}        & \multicolumn{3}{c}{Optical Depths$\cnote$}                & \multicolumn{6}{c}{Absolute Relative Difference in $\Teq(\theta)$}                                                \\
                                    &                                   &                           &                                   &               &                               &           & \multicolumn{2}{c}{Multigroup$\dnote$}&\multicolumn{2}{c}{Hybrid$\enote$} &\multicolumn{2}{c}{RADMC-3D$\fnote$}   \\
                                    &                                   &                           &                                   &               &                               &           & Avg.      & Max.                      & Avg.      & Max.                  & Avg.      & Max.                      \\
        $(\cm^2\gm^{-1})$           &                                   &                           & \multicolumn{1}{c}{$(\degree)$}   &               &                               &           & (\%)      & (\%)                      & (\%)      & (\%)                  & (\%)      & (\%)                      \\
        \multicolumn{1}{c}{(1)}     & \multicolumn{1}{c}{(2)}           & \multicolumn{1}{c}{(3)}   & \multicolumn{1}{c}{(4)}           & \multicolumn{3}{c}{(5)}                                   & (6)       & (7)                       & (8)       & (9)                   & (10)      & (11)}
    \startdata
        $\kabs = 300$               & \ref{sec:gray_with_dust_emission} & \ref{fig:gray_absorption} & 30                                & $10^{-11}$    & $\lesssim \tau \lesssim$      & $10^{-8}$ & 1         & 3                         & $<1$      & 1                     & \nodata   & \nodata                   \\
                                    &                                   &                           & 10                                & $10^{-7}$     & $\lesssim \tau \lesssim$      & $1$       & 1         & 1                         & 1         & 2                     & \nodata   & \nodata                   \\
                                    &                                   &                           & 0                                 & $10^{-3}$     & $\lesssim \tau \lesssim$      & $10^2$    & 5         & 11                        & 6         & 16                    & \nodata   & \nodata                   \\
        $\knuabs$                   & \ref{sec:with_dust_emission}      & \ref{fig:absorption}      & 30                                & $10^{-10}$    & $\lesssim \tau_* \lesssim$    & $10^{-7}$ & 2         & 3                         & 1         & 2                     & \nodata   & \nodata                   \\
                                    &                                   &                           & 10                                & $10^{-6}$     & $\lesssim \tau_* \lesssim$    & $10$      & 2         & 4                         & 8         & 32                    & \nodata   & \nodata                   \\
                                    &                                   &                           & 0                                 & $10^{-2}$     & $\lesssim \tau_* \lesssim$    & $10^3$    & 5         & 8                         & 13        & 18                    & \nodata   & \nodata                   \\
        $\knuabssca$                & \ref{sec:scattering}              & \ref{fig:scattering}      & 30                                & $10^{-9}$     & $\lesssim \tau_* \lesssim$    & $10^{-6}$ & 3         & 4                         & \nodata   & \nodata               & \nodata   & \nodata                   \\
                                    &                                   &                           & 10                                & $10^{-5}$     & $\lesssim \tau_* \lesssim$    & $10^2$    & 2         & 4                         & \nodata   & \nodata               & \nodata   & \nodata                   \\
                                    &                                   &                           & 0                                 & $10^{-1}$     & $\lesssim \tau_* \lesssim$    & $10^4$    & 5         & 8                         & \nodata   & \nodata               & \nodata   & \nodata                   \\
        $\knuabsscag$               & \ref{sec:scattering}              & \ref{fig:scattering_modes}& 30                                & $10^{-9}$     & $\lesssim \tau_* \lesssim$    & $10^{-6}$ & \nodata   & \nodata                   & \nodata   & \nodata               & 1         & 1                         \\
                                    &                                   &                           & 10                                & $10^{-5}$     & $\lesssim \tau_* \lesssim$    & $10^2$    & \nodata   & \nodata                   & \nodata   & \nodata               & 1         & 6                         \\
                                    &                                   &                           & 0                                 & $10^{-1}$     & $\lesssim \tau_* \lesssim$    & $10^4$    & \nodata   & \nodata                   & \nodata   & \nodata               & 3         & 9                         \\
    \enddata
    \tablecomments{Columns give the (1) opacity profile(s);$\anote$ corresponding (2) section and (3) figure detailing the full comparison; (4) meridional latitude of the radial profile $\Teq(\theta)$;$\bnote$ (5) range in (effective) optical depth$\gnote$ $\tau$ ($\tau_*$) to (peak) stellar irradiation;$\cnote$ (6) average (arithmetic mean) and (7) maximum absolute relative difference in $\Teq(\theta)$ between our multigroup$\dnote$ hydrostatic model$\hnote$ with $N_f = 64$ frequency bands and RADMC-3D;$\fnote$ (8) average and (9) maximum between the hydrostatic hybrid$\enote$ model with gray thermal reemission and RADMC-3D; and (10) average and (11) maximum between RADMC-3D models with isotropic $\knuabssca$ or anisotropic $\knuabsscag$ scattering opacities.\\
        $\anote$ Figure~\ref{fig:opacities}\\
        $\bnote$ Figure~\ref{fig:thermal_equilibrium}\\
        $\cnote$ Figure~\ref{fig:disk}\\
        $\dnote$ Section~\ref{sec:methodology}\\
        $\enote$ Section~\ref{sec:hybrid_ray_tracing}\\
        $\fnote$ Section~\ref{sec:radmc-3d}\\
        $\gnote$ Equation~\eqref{eq:tau_*}\\
        $\hnote$ Section~\ref{sec:model}}
\end{deluxetable*}

\subsubsection{With Dust Emission} \label{sec:gray_with_dust_emission}

In numerical models where the dust absorbs energy from stellar irradiation, the dust will heat up until it reaches thermal equilibrium (Section~\ref{sec:thermal_equilibrium}).
We compare equilibrium temperature solutions from RADMC-3D (Section~\ref{sec:radmc-3d}) and Athena++ which include our multigroup radiation transport framework (Section~\ref{sec:methodology}) and a hybrid ray-tracing method for stellar irradiation (Section~\ref{sec:hybrid_ray_tracing}).
For this comparison, we adopt a gray absorption opacity $\kabs = 300\cm^2\gm^{-1}$ for $N_f = 1$ in Athena++ and for each point in the RADMC-3D wavelength grid.

\begin{figure*}
    \includegraphics[width=\textwidth]{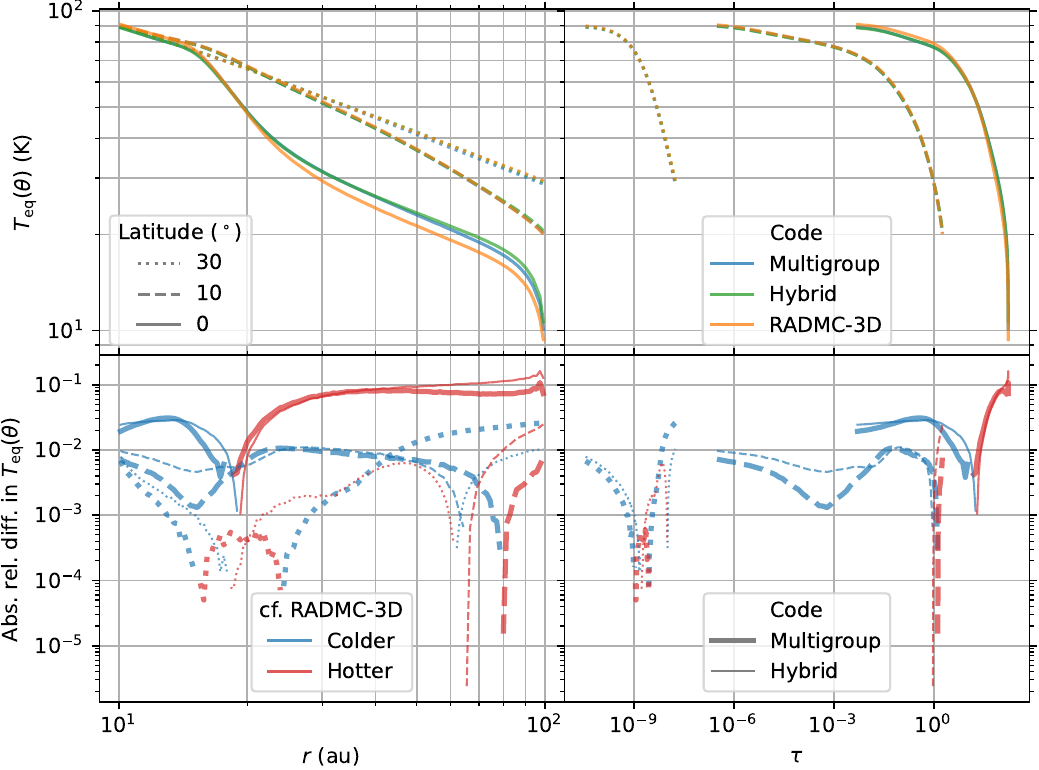}
    \caption{Equilibrium temperatures $\Teq$ from models sharing the same gray absorption opacity $\kabs = 300\cm^2\gm^{-1}$.
    The left panels show radial profiles, while the right ones show them as functions of gray optical depth $\tau$ [equation~\eqref{eq:formal_transfer_solution}].
    The upper panels show absolute temperatures, while the lower ones show the absolute relative (i.e., fractional) difference between RADMC-3D (Section~\ref{sec:radmc-3d}) and Athena++ with either our multigroup radiation transport framework (Section~\ref{sec:methodology}) or a hybrid ray-tracing method for stellar irradiation (Section~\ref{sec:hybrid_ray_tracing}).
    Lines styles correspond to profiles at different disk latitudes (cf. Figure~\ref{fig:thermal_equilibrium}).
    Following their respective legends, colors show different codes in the upper panels or show where the relative difference is negative (blue) or positive (red)---i.e., below (colder) or above (hotter) than RADMC-3D---in the lower panels, where line weights instead differentiate the codes.}
    \label{fig:gray_absorption}
\end{figure*}

The upper panels of Figure~\ref{fig:gray_absorption} compares the equilibrium temperature profiles $\Teq$ reached by the three methods at the same latitudes in Figure~\ref{fig:attenuation}.
The upper-left panel shows the disk atmosphere temperature at $30\degree$ reaching $90\K$ at $\rmin$ with $\Teq \propto r^{-1/2}$, consistent with an analytic approximation by \cite{Spitzer1978} for distantly irradiated dust in the gray and isotropic limit \citep[cf.][eq.~4]{KuiperKlessen2013}.
Moreover, the optically thin region near $\rmin$, afforded by the softened radial density profile (Figure~\ref{fig:midplane}), allows the midplane ($0\degree$) and intermediate profiles ($10\degree$) to match the atmosphere for $\Teq(r < 15\au)$.
However, $\Teq(r > 30\au)$ at the midplane tracks $20\K$ below that at $30\degree$, with that at $10\degree$ gradually cooling to $10\K$ below the atmosphere by $\rmax$.
Between the solutions, $\Teq$ is slightly higher in Athena++ than RADMC-3D in optically thick regions (e.g., for $r > 20\au$ along the midplane), consistent with the analytic flux comparisons in Section~\ref{sec:without_dust_emission} and Figure~\ref{fig:attenuation}.

The lower panels of Figure~\ref{fig:gray_absorption} compare the absolute relative (i.e., fractional) difference in $\Teq$ between each of the Athena++ models and RADMC-3D, with the averages and maxima listed in Table~\ref{tab:hydrostatic_comparisons}.
As a function of $\tau$, the lower-right panel shows Athena++ agrees with RADMC-3D to within 1\% on average in the optically thin atmosphere and at intermediate optical depths ($10\degree$).
At the optically thick midplane, our multigroup framework and hybrid ray tracing agree on average to within 5\% and 6\% of RADMC-3D, respectively, and  at worst to within 11\% and 16\%, respectively.
Consistent with both under- and overestimating the expected analytic flux (Section~\ref{sec:without_dust_emission}), both Athena++ methods tend to be slightly colder and hotter than RADMC-3D in relatively optically thin and thick regions, respectively, at each meridional latitude.
Lastly, the sharp drops by many orders of magnitude in the lower panels (e.g., at $\tau \approx 1$ and 10) correspond to where Athena++ $\Teq$ profiles in the upper panels transition from under- to overestimating those from RADMC-3D.

\subsection{Frequency-dependent Opacities} \label{sec:frequency-dependent_opacities}

We compare models in which the opacities depend on frequency.
For Athena++ models with our self-consistent multigroup radiation transport framework (Section~\ref{sec:methodology}), we use $N_f = 64$ frequency bands (Section~\ref{sec:radiation_transport}) with their corresponding band-mean opacities (Section~\ref{sec:emissivity_opacity}) from dust with a DSHARP composition (Section~\ref{sec:dust_opacities}).
Section~\ref{sec:with_dust_emission} compares equilibrium temperature solutions that only include frequency-dependent absorption, while Section~\ref{sec:scattering} compares those that also include scattering.

\subsubsection{With Dust Emission} \label{sec:with_dust_emission}

Similar to Section~\ref{sec:gray_with_dust_emission} but for models with frequency-dependent absorption opacities, we compare equilibrium temperatures $\Teq$ from RADMC-3D (Section~\ref{sec:radmc-3d}) and Athena++ which include our multigroup radiation transport framework (Section~\ref{sec:methodology}) and a hybrid ray-tracing method for stellar irradiation (Section~\ref{sec:hybrid_ray_tracing}).
In the hybrid model, the ray-traced stellar irradiation uses the same monochromatic absorption opacity $\knuabs$ (Section~\ref{sec:dust_opacities}) we use in RADMC-3D to also account for dust heating.

\begin{figure*}
    \includegraphics[width=\textwidth]{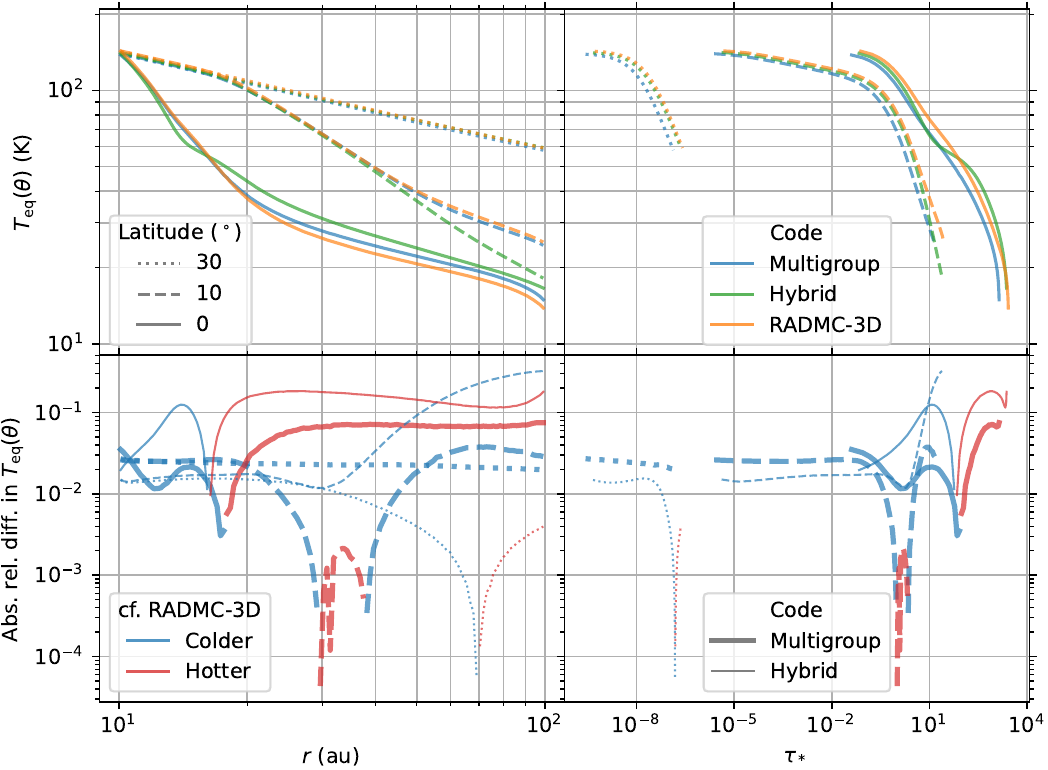}
    \caption{Similar to Figure~\ref{fig:gray_absorption} except for models using the frequency-dependent absorption opacities $\knuabs$ shown in Figure~\ref{fig:opacities}.
    The right panels show the profiles as functions of optical depth to peak stellar irradiation $\tau_*$ [equation~\eqref{eq:tau_*}].
    The Athena++ model with our multigroup radiation transport framework (Section~\ref{sec:methodology}) uses $N_f = 64$ logarithmically uniform frequency bands, the hybrid Athena++ model with frequency-dependent ray-traced stellar irradiation uses gray thermal reemission (Section~\ref{sec:hybrid_ray_tracing}), and the RADMC-3D model uses 150 wavelength points (Section~\ref{sec:radmc-3d}).}
    \label{fig:absorption}
\end{figure*}

Similar to Figure~\ref{fig:gray_absorption}, Figure~\ref{fig:absorption} compares $\Teq$ profiles at the same meridional latitudes.
Compared with the gray case, $\max(\Teq) \approx 150\K$, as these frequency-dependent dust opacities preferentially absorb ultraviolet stellar irradiation [cf. $\knuabs$ and $B_\nu(T_*)$ in Figure~\ref{fig:opacities}].
For the midplane in the upper-left panel, $\od\Teq/\od r$ is much steeper for $r < 20\au$, but $\Teq(r > 30\au$) follows a similar absolute profile despite being about $60\K$ colder than at $30\degree$ latitude.
This difference between atmospheric and midplane temperatures is three times greater than in gray models (cf. Figure~\ref{fig:gray_absorption}, upper-left panel).
At $10\degree$, $\od\Teq/\od r$ is steeper for $20 < r\au < 50$.
However, this profile from hybrid ray tracing continues to cool toward the midplane value at $\rmax$, while those from the other two methods stay roughly $10$ to $20\K$ warmer for $r > 50\au$.

Figure~\ref{fig:absorption} also shows the absolute $\Teq$ as a function of effective depth to peak stellar irradiation $\tau_*$ in the upper-right panel. 
Since the opacities to stellar irradiation for the Athena++ hybrid ray-tracing and RADMC-3D models share the same frequency resolution, we can use the same $\kstarabs$ to calculate $\tau_*$ via equation~\eqref{eq:tau_*} for these two methods, resulting in their profiles sharing the exact same domains in $\tau_*$ at each latitude.
However, in our multigroup radiation transport framework, $f_*$ and the mean opacities at that band only approximate $\nu_*$ and $\kstarabs$, respectively (Section~\ref{sec:dust_opacities}), according to equation~\eqref{eq:f_*-to-nu_*_limit}.
Thus, compared to hybrid ray tracing and RADMC-3D, the domains in $\tau_*$ for these profiles displace a bit to the left due to a slight under approximation with $N_f = 64$ bands (cf. Section~\ref{sec:frequency_bands} and Figure~\ref{fig:nfreq}).

For the absolute relative differences between each Athena++ model and that from RADMC-3D in the panel directly below (lower right), we use the same $\kstarabs$ which aligns these domains with those from RADMC-3D above.
Compared to their purely gray counterparts (Figure~\ref{fig:gray_absorption}), hybrid ray tracing differs from RADMC-3D by more than 30\% at $10\degree$ (Table~\ref{tab:hydrostatic_comparisons}) toward $\rmax$.
We attribute this marked difference to the fact that dust reemission and subsequent thermalization are frequency independent under the hybrid scheme (Section~\ref{sec:hybrid_ray_tracing} and Appendix~\ref{appx:hybrid_ray-tracing_method}), and thus the disk does not cool as accurately, especially at intermediate latitudes.
Conversely, our multigroup framework differs by at most 8\% along the midplane ($\tau_* \sim 10^3$) and 4\% at $10\degree$ toward $\rmax$ ($\tau_* \lesssim 10$).
Finally, although both methods can be colder and hotter than RADMC-3D (consistent with our analytic comparisons in Section~\ref{sec:without_dust_emission}), we note that this multigroup model remains cooler throughout the entire disk atmosphere, i.e., at $10\degree$ or for $\tau_* \lesssim 1$.

\subsubsection{Scattering} \label{sec:scattering}

Similar to Section~\ref{sec:with_dust_emission} but for models that also include frequency-dependent isotropic scattering opacities, we compare equilibrium temperatures $\Teq$ from our multigroup radiation transport framework in Athena++ (Section~\ref{sec:methodology}) and RADMC-3D (Section~\ref{sec:radmc-3d}).
We include the monochromatic scattering opacity $\knusca$ (gray curve in the right panel of Figure~\ref{fig:opacities}) in our RADMC-3D model and the corresponding band-mean opacities for $N_f = 64$ (Section~\ref{sec:dust_opacities}) in our multigroup Athena++ model.

\begin{figure*}
    \includegraphics[width=\textwidth]{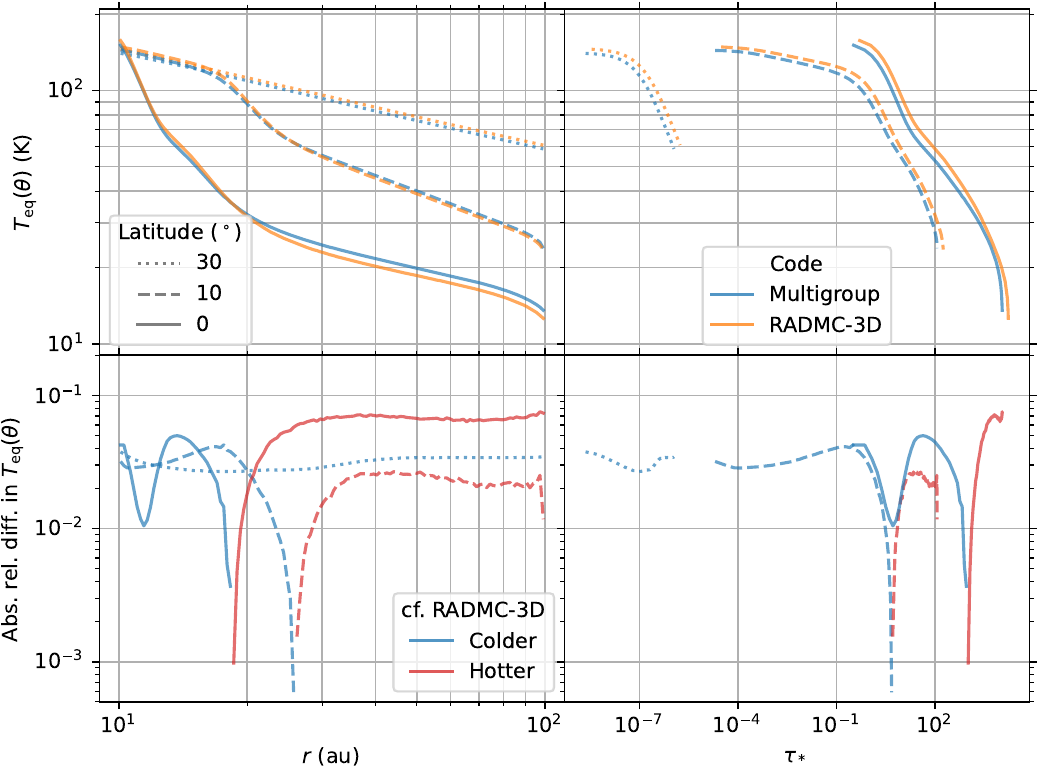}
    \caption{Similar to Figure~\ref{fig:absorption} except for models that also include the frequency-dependent isotropic scattering opacities $\knusca$ (Section~\ref{sec:scattering}) shown in Figure~\ref{fig:opacities}.}
    \label{fig:scattering}
\end{figure*}

Similar to Figures~\ref{fig:gray_absorption} and \ref{fig:absorption}, Figure~\ref{fig:scattering} compares $\Teq$ profiles at the same latitudes.
As in the case of only frequency-dependent absorption, $\max(\Teq) \approx 150\K$, since scattering has little effect on temperatures in effectively thin regions, e.g., at $30\degree$.
However, along the midplane, $\od\Teq/\od r$ for $r < 20\au$ is even steeper (especially near $\rmin$) in the upper-left panel compared to that in Figure~\ref{fig:absorption} as incoming stellar irradiation is increasingly scattered out of the radial direction, leaving the profile up to $10\K$ cooler at $20\au$ than without scattering.
At $10\degree$, $\od\Teq/\od r$ is steeper around $r \approx 20\au$ but follows a profile similar to that without scattering for $\Teq(r > 50\au)$.
Lastly, similar levels of agreement with RADMC-3D are maintained (cf. lower panels), differing again by up to 8\% and 4\% at the midplane and higher latitudes, respectively (Table~\ref{tab:hydrostatic_comparisons}), while remaining consistently cooler for $\tau_* \lesssim 1$ (cf. Section~\ref{sec:with_dust_emission}).

Since the Athena++ radiation transport module (Section~\ref{sec:radiation_transport}) does not currently support anisotropic scattering (Section~\ref{sec:emissivity_opacity}), we investigate its effect on stellar-irradiated protoplanetary disk temperatures by solely using RADMC-3D.
Qualitatively, it redistributes stellar photons in both direction and path length, which alters where energy is ultimately deposited in the disk; e.g., forward‑peaked (non‑isotropic) scattering allows radiation to penetrate deeper before being absorbed, thereby modifying the vertical heating profile and the resulting temperature structure.
One simplified approach to modeling anisotropic scattering uses the \cite{HenyeyGreenstein1941} parameter $g = \langle\cos(\theta)\rangle$, where $\theta$ is the scattering angle in this context.
We use the same Python package introduced in Section~\ref{sec:dust_opacities} \citep{BirnstielDullemondZhu2018} to calculate a monochromatic $g_\nu$ via Mie scattering for our DSHARP grain-size distribution [equation~\eqref{eq:grain-size_distribution}] to use in RADMC-3D.

\begin{figure*}
    \includegraphics[width=\textwidth]{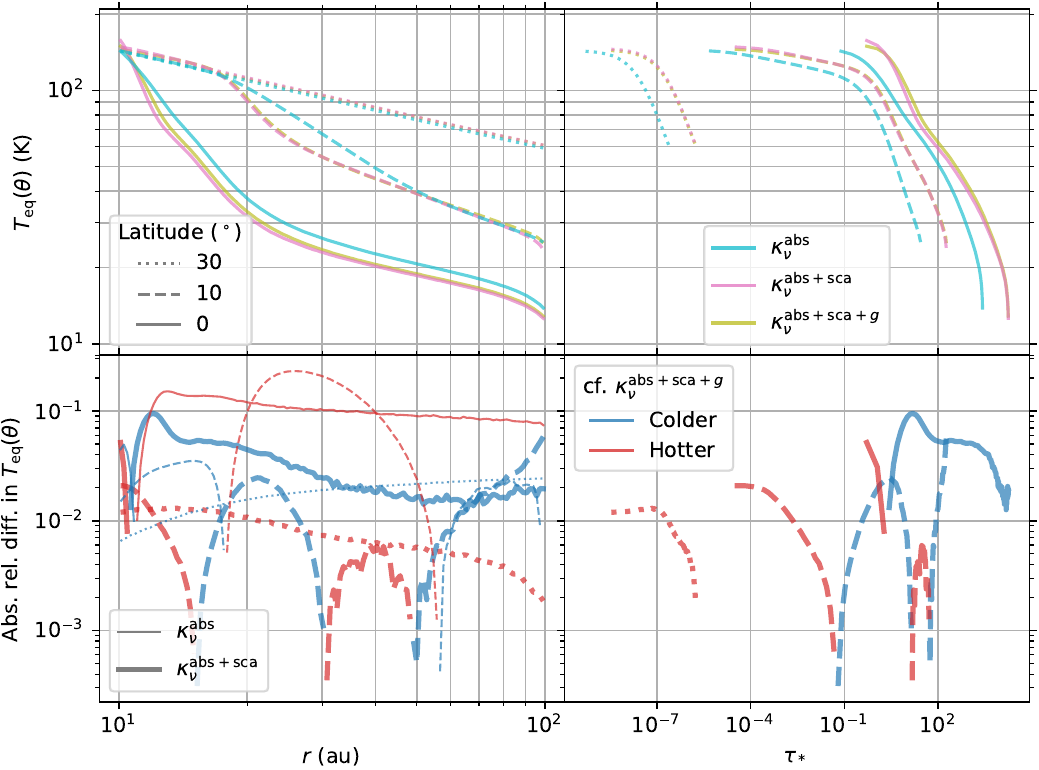}
    \caption{Similar to Figure~\ref{fig:scattering} except for different scattering modes only in RADMC-3D.
    In upper panels, colors show models with only absorption opacities $\knuabs$ (cyan), with
    both absorption and isotropic scattering opacities $\knuabssca$ (pink),
    and with both absorption and anisotropic scattering opacities $\knuabsscag$ (olive; Section~\ref{sec:scattering}).
    In the lower panels, line weights show the absolute relative difference of those models against that with both absorption and anisotropic scattering opacities $\knuabsscag$, where blue or red colors show where those models are colder or hotter, respectively, than that with $\knuabsscag$.
    As the domains in effective depth to peak stellar irradiation $\tau_*$ between models with and without scattering differ by an order of magnitude (upper-right panel), the lower-right panel shows the absolute relative difference only between models with scattering.}
    \label{fig:scattering_modes}
\end{figure*}

Figure~\ref{fig:scattering_modes} compares $\Teq$ profiles only from RADMC-3D for opacities that (1) only include monochromatic absorption $\kappa_\nu^\mathrm{abs}$, (2) also include isotropic scattering $\kappa_\nu^\mathrm{abs+sca}$, or (3) also include anisotropic scattering $\kappa_\nu^{\mathrm{abs+sca}+g}$.
In the upper-right panel, the domains in $\tau_*$ differ by an order of magnitude between models with or without scattering (cf. $\kstarabs$ and $\kstarsca$ in the right panel of Figure~\ref{fig:opacities}).
We only plot the absolute relative difference between the isotropic and anisotropic scattering models in the panel directly below (lower right) which is 3\% or less on average (Table~\ref{tab:hydrostatic_comparisons}) but at most 9\% along the midplane ($\tau_* \approx 10$) and 6\% at $10\degree$ ($\tau_* \approx 10^2$).
Compared to isotropic scattering in the lower panels, anisotropic scattering tends to make the disk midplane and atmosphere only slightly warmer and cooler, respectively.

\subsection{Resolution Studies} \label{sec:resolution_studies}

We compare additional models from our multigroup framework (Section~\ref{sec:methodology}) by varying key resolution parameters in the Athena++ radiation transport module.
As in Section~\ref{sec:scattering}, all models in this section include frequency-dependent absorption and isotropic scattering opacities as specified in Section~\ref{sec:dust_opacities}.
Sections~\ref{sec:frequency_bands} and \ref{sec:angles} study the convergence of these models as functions of the numbers of frequency bands or cell-centered polar angles, respectively, and Section~\ref{sec:performance} compares their computational performance.

\begin{deluxetable*}{lcrcrrrrrrr}
    \tablecaption{Resolution Studies\label{tab:resolution_studies}}
    \tablecolumns{11}
    \tablehead{
        \colhead{Section}           & \colhead{Figure}              & \colhead{$N_f$}         & \colhead{$N_\zeta$}       & \multicolumn{6}{c}{Absolute Relative Difference in $\Teq(\theta)$}                                                                                                                                & \colhead{Core Hours$\anote$}    \\
                                    &                               &                         &                           & \multicolumn{2}{c}{Atmosphere}                                    & \multicolumn{2}{c}{Intermediate}                              & \multicolumn{2}{c}{Midplane}                                  &                                 \\
                                    &                               &                         &                           & \multicolumn{2}{c}{$30\degree$}                                   & \multicolumn{2}{c}{$10\degree$}                               & \multicolumn{2}{c}{$0\degree$}                                &                                 \\
                                    &                               &                         &                           & \multicolumn{2}{c}{$10^{-9} \lesssim \tau_* \lesssim 10^{-6}$}    & \multicolumn{2}{c}{$10^{-5} \lesssim \tau_* \lesssim 10^2$}   & \multicolumn{2}{c}{$10^{-1} \lesssim \tau_* \lesssim 10^4$}   &                                 \\
                                    &                               &                         &                           & Avg.                          & Max.                              & Avg.                          & Max.                          & Avg.                          & Max.                          &                                 \\
                                    &                               &                         &                           & (\%)                          & (\%)                              & (\%)                          & (\%)                          & (\%)                          & (\%)                          &                                 \\
        \multicolumn{1}{c}{(1)}     & (2)                           & \multicolumn{1}{c}{(3)} & \multicolumn{1}{c}{(4)}   & (5)                           & (6)                               & (7)                           & (8)                           & (9)                           & (10)                          & \multicolumn{1}{c}{(11)}}
    \startdata
        \ref{sec:performance}       & \ref{fig:performance}         & \nodata                 & \nodata                   & \nodata                       & \nodata                           & \nodata                       & \nodata                       & \nodata                       & \nodata                       &   0.03                          \\
        \ref{sec:optimization}      & \ref{fig:optimized_profiles}  & 3                       & 4                         & 9                             & 9                                 & 7                             & 19                            & 11                            & 17                            &   8.70                          \\
        \ref{sec:frequency_bands}   & \ref{fig:nfreq}               & 4                       & 4                         & 26                            & 27                                & 11                            & 28                            & 10                            & 33                            &   9.46                          \\
                                    &                               & 8                       & 4                         & 6                             & 7                                 & 3                             & 13                            & 9                             & 13                            &  22.79                          \\
                                    &                               & 16                      & 4                         & 8                             & 9                                 & 4                             & 9                             & 8                             & 15                            &  25.20                          \\
                                    &                               & 32                      & 4                         & 6                             & 6                                 & 3                             & 6                             & 6                             & 9                             &  41.47                          \\
        \ref{sec:angles}            & \ref{fig:nzeta}               & 64                      & 1                         & 3                             & 4                                 & 6                             & 9                             & 13                            & 20                            &  39.52                          \\
                                    &                               &                         & 2                         & 3                             & 4                                 & 3                             & 5                             & 8                             & 11                            &  52.28                          \\
                                    &                               &                         & 4                         & 3                             & 4                                 & 2                             & 4                             & 5                             & 8                             &  85.17                          \\
                                    &                               &                         & 8                         & 3                             & 4                                 & 2                             & 4                             & 4                             & 6                             & 159.67                          \\
    \enddata
    \tablecomments{Columns give the corresponding (1) section and (2) figure detailing the study; (3) logarithmically uniform frequency bands $N_f$ (except for $N_f = 3)\bnote$ and (4) cell-centered polar angles $N_\zeta$ between 0 and $\pi/2$;$\cnote$ (5) average (arithmetic mean) and (6) maximum absolute relative difference in radial profiles of $\Teq(\theta)$ in the disk atmosphere at a meridional latitude of $30\degree$ with a $10^{-9} \lesssim \tau_* \lesssim 10^{-6}$ range in effective depth to peak stellar irradiation;$\dnote$ the (7) average and (8) maximum at an intermediate latitude of $10\degree$ with $10^{-5} \lesssim \tau_* \lesssim 10^2$; and the (9) average and (10) maximum in the disk midplane at $0\degree$ with $10^{-1} \lesssim \tau_* \lesssim 10^4$; and (11) total core hours (i.e., the multiplicative product of the number of CPU cores and the elapsed real time) used in hours to integrate up to the same simulation time limit.$\anote$
    Except for the one without radiation in the first row, all models listed here use the frequency-dependent dust absorption and scattering opacities detailed in Section~\ref{sec:dust_opacities} and shown in Figure~\ref{fig:opacities}.\\
        $\anote$ Section~\ref{sec:performance} and Figure~\ref{fig:performance}\\
        $\bnote$ Nonuniform, custom bands (Section~\ref{sec:optimization} and Figure~\ref{fig:optimized_opacities})\\
        $\cnote$ Section~\ref{sec:radiation_transport} and Figures~\ref{fig:nhat} and \ref{fig:opacities}\\
        $\dnote$ Equation~\eqref{eq:tau_*} and Figure~\ref{fig:disk}}
\end{deluxetable*}

\subsubsection{Frequency Bands} \label{sec:frequency_bands}

To assess the convergence implied by equations~\eqref{eq:kf-to-knu_limit} and \eqref{eq:f_*-to-nu_*_limit}, we continue to double the number of frequency bands from $N_f = 4$ to 64, while maintaining the same angular resolution $N_\zeta = 4$ (Table~\ref{tab:mathematical_notation}; cf. Section~\ref{sec:angles}), and compare the equilibrium temperatures $\Teq$ (Section~\ref{sec:thermal_equilibrium}).
Except for the first $f = 0 = [0, \nu_1)$, and the last $f = N_f - 1 = [\nu_{N_f-1}, \infty)$, the bands are logarithmically uniform with $\nu_1$ and $\nu_{N_f-1}$ chosen to best cover the $3\times10^{9}\Hz < \nu < 3\times10^{15}\Hz$ range of monochromatic opacities specified in Section~\ref{sec:dust_opacities} (cf. $N_f = 4$ and $\knuabs$ in the right panel of Figure~\ref{fig:opacities}).

\begin{figure*}
    \includegraphics[width=\textwidth]{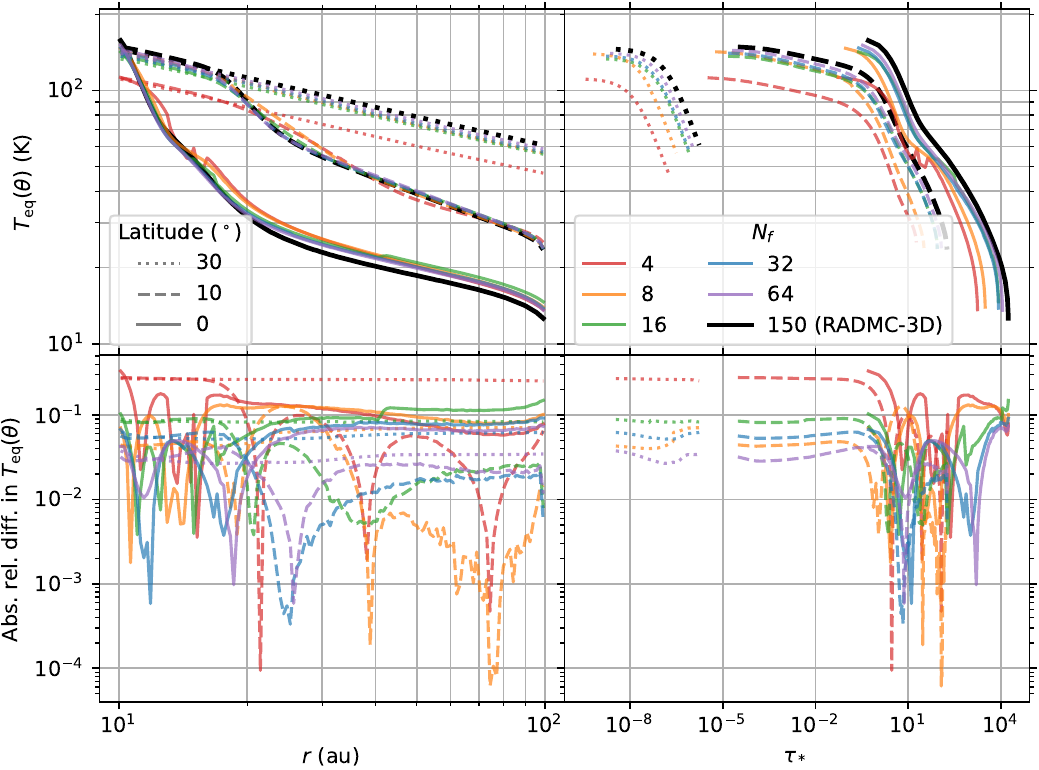}
    \caption{Similar to Figure~\ref{fig:scattering} except for a range in the number of frequency bands $N_f$.
    Colored curves show Athena++ models with different logarithmically uniform $N_f$ (Section~\ref{sec:frequency_bands}), while black ones show the RADMC-3D model with 150 wavelength points (Section~\ref{sec:radmc-3d}).
    The lower panels show the absolute relative difference between each Athena++ model and RADMC-3D.}
    \label{fig:nfreq}
\end{figure*}

Like Figures~\ref{fig:gray_absorption}--\ref{fig:scattering_modes}, Figure~\ref{fig:nfreq} compares $\Teq$ of each $N_f$ model from Athena++ and RADMC-3D with 150 wavelength points (Section~\ref{sec:radmc-3d}).
In general, $\max(\Teq)$ and the absolute scale of $\Teq(r)$ at $30\degree$ latitude in the upper panels increase with $N_f$, since the mean opacities in additional bands higher than $\nu_*$ allow effectively thin media to absorb additional energy from the ultraviolet end of stellar irradiation [cf. $\kpf(T_*)$, $\krf(T_*)$, and $B_\nu(T_*)$ in the right panel of Figure~\ref{fig:opacities}].
Moreover, these trends converge toward the solution from RADMC-3D.

Figure~\ref{fig:nfreq} also compares $\Teq$ as a function of the effective depth to peak stellar irradiation $\tau_*$ [equation~\eqref{eq:tau_*}] in the upper-right panel.
As explained in Section~\ref{sec:with_dust_emission} and also shown in the upper-right panels of Figures~\ref{fig:absorption} and \ref{fig:scattering}, the approximation of $\kstarabs$ and $\kstarsca$ by the mean opacities at $f_*$ displaces the domains of $\tau_*$ depending on $N_f$.
Although these domains tend to converge to those from RADMC-3D as $N_f$ increases, the magnitudes of these displacements are not strictly monotonic (cf. $N_f = 16$ and 32).
As a result of logarithmically uniform spacing, the band interface closest to the sharp bump in $\knuabs$ at $10^{14}\Hz$ (Figure~\ref{fig:opacities}, right panel; Section~\ref{sec:dust_opacities}) affects the magnitude of the mean opacities in neighboring bands [Section~\ref{sec:emissivity_opacity} and equations~\eqref{eq:kpfe}--\eqref{eq:ksf}].
Given their close proximity in frequency to the peak stellar irradiation at $\nu_* \approx 3.40\times10^{14}\Hz$ [Section~\ref{sec:stellar-irradiated_disk} and $B_\nu(T_*)$ in the right panel of Figure~\ref{fig:opacities}], this ultimately affects the mean opacities at $f_*$ which we use to calculate $\tau_*$ via equation~\eqref{eq:tau_*}.

As in Figures~\ref{fig:absorption} and \ref{fig:scattering}, the lower panels of Figure~\ref{fig:nfreq} compare the absolute relative differences in $\Teq$ between each Athena++ model and that from RADMC-3D with averages and maxima listed in Table~\ref{tab:resolution_studies}.
We use the same $\kstarabs + \kstarsca$ to align the domains in $\tau_*$ only for the lower-right panel (cf. Section~\ref{sec:with_dust_emission}) to help compare the range in maximum absolute relative differences from 33\% for $N_f = 4$ to 8\% for $N_f = 64$.
For $\tau_* < 1$ (e.g., at latitudes of $10\degree$ and $30\degree$), the absolute relative differences decrease by an order of magnitude from $N_f = 4$ to 64 but their averages do not monotonically decrease as $N_f$ increases, with those for $N_f = 8$ lower than those for 16 bands.
Similar to the non-monotonic convergence of the domains in $\tau_*$ in the panel directly above, this trend is an approximation artifact of how the additional mean opacities at $f \geq f_*$, which absorb $B_\nu(T_*)$ at $\nu \gtrapprox \nu_*$ and primarily heat the dust in effectively thin regions, do not strictly increase with $N_f$ depending on the positions and spacing of the bands around the $\knuabs$ bump at $10^{14}\Hz$.
Nevertheless, as shown by Table~\ref{tab:resolution_studies}, the overall accuracy improves as $N_f$ approaches the number frequency points used in RADMC-3D, since the resulting approximation error in $\kstarabs + \kstarsca$ dominates the inaccuracy of the extinction coefficient at low densities [equation~\eqref{eq:tau_*}].
On the other hand, since high (midplane) densities dominate $\tau_* \sim 10^4$, approximating $\kstarabs + \kstarsca$ results in at most a 50\% difference between $N_f$.

\subsubsection{Angles} \label{sec:angles}

As introduced in Section~\ref{sec:radiation_transport} and visualized in Figure~\ref{fig:nhat}, $N_\zeta$ is the number of cell-centered polar angles between 0 and $\pi/2$.
For axisymmetric models, $N_\zeta$ alone affects the degree to which (re)emission and scattering can be transported isotropically.
Thus, according to equation~\eqref{eq:w_n-to-dOmega_limit}, increasing $N_\zeta$ (and $N_\psi$) should result in more accurate (3D) models and better agreement with Monte Carlo radiative transfer methods, including RADMC-3D (Section~\ref{sec:radmc-3d}), where photon reemission and isotropic scattering are randomly drawn from a uniform angular distribution.

\begin{figure*}
    \includegraphics[width=\textwidth]{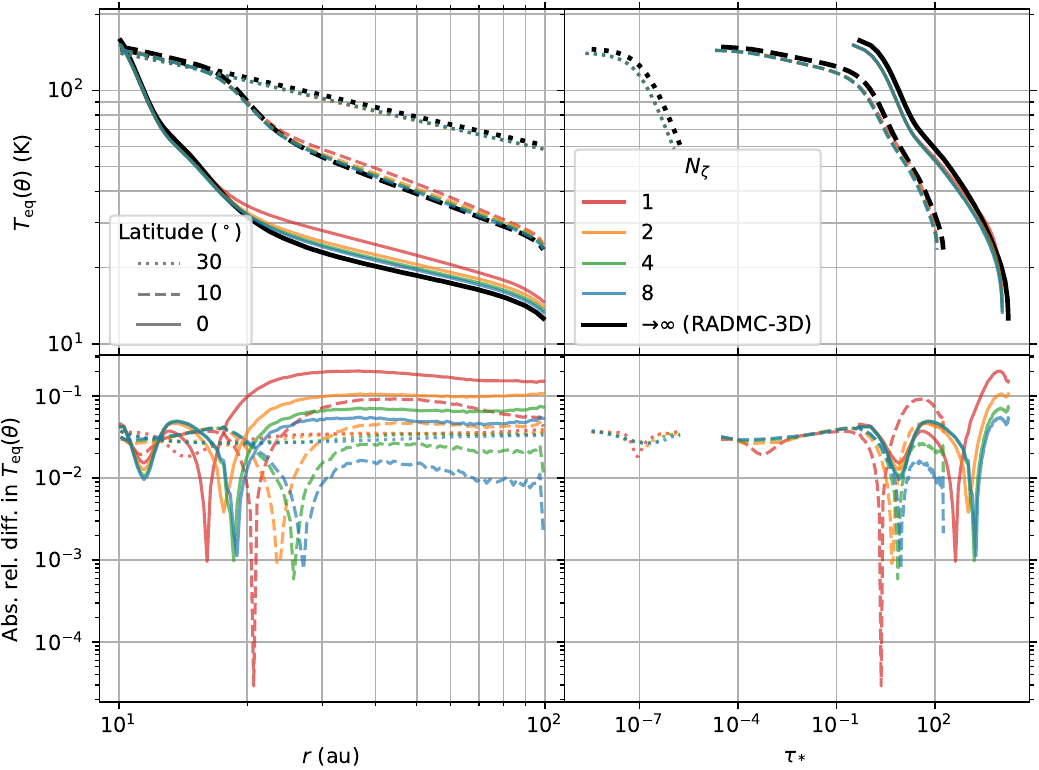}
    \caption{Similar to Figure~\ref{fig:scattering} except for a range in the number of cell-centered polar angles $N_\zeta$ between 0 and $\pi/2$ (Section~\ref{sec:radiation_transport}).
    Colored curves show Athena++ models with different $N_\zeta$, while black ones show the RADMC-3D model where isotropic photon reemission and scattering angles are randomly drawn from a uniform distribution.
    The lower panels show the absolute relative difference of each Athena++ model with RADMC-3D.}
    \label{fig:nzeta}
\end{figure*}

To assess the convergence implied by equation~\eqref{eq:w_n-to-dOmega_limit}, we continue to double $N_\zeta$ from 1 to 8 while keeping $N_f = 64$.
Like Figure~\ref{fig:nfreq}, Figure~\ref{fig:nzeta} compares $\Teq$ of each $N_\zeta$ model from Athena++ and RADMC-3D, where the effective angular resolution in the latter increases with the number of photon packets (Section~\ref{sec:radmc-3d}).
In Sections~\ref{sec:with_dust_emission} and \ref{sec:frequency_bands}, the systematic difference between Athena++ and RADMC-3D, with respect to $\max(\Teq)$ (top panels) and the domains of effective optical depth $\tau_*$ (upper-right panel), is due to the approximations of the band-mean opacities to $\knuabs + \knusca$ (Section~\ref{sec:emissivity_opacity}) and of $f_*$ to $\nu_*$ (Section~\ref{sec:dust_opacities}), as encapsulated by equations~\eqref{eq:kf-to-knu_limit} and \eqref{eq:f_*-to-nu_*_limit}, respectively.

As revealed by Figure~\ref{fig:nzeta} and Table~\ref{tab:resolution_studies}, convergence and agreement with RADMC-3D as a function of $N_\zeta$ depend on $\tau_*$.
For $\tau_* < 1$ (e.g., $r < 20\au$ and at $30\degree$ latitude), Athena++ models are almost invariant as a function of $N_\zeta$, as the radial profile tends to agree with purely radial ray tracing in the effectively thin limit [cf. the analytic solution by equation~\eqref{eq:F_r-tau}].
However, at high optical depths (e.g., at $0\degree$ and $10\degree$), $\Teq(r > 20\au)$ lowers overall as $N_\zeta$ increases, as the degree to which $N_\zeta$ can isotropically scatter and transport (re)emission (away from the radial direction due to angular equipartition) helps to shield midplane regions from stellar irradiation and to enhance radiative cooling of effectively thick media, respectively.
Moreover, the absolute relative differences with RADMC-3D in the lower panels and in Table~\ref{tab:resolution_studies} show a more monotonic trend with $N_\zeta$, unlike the frequency resolution study (Section~\ref{sec:frequency_bands}), with a maximum difference of 6\% for $N_\zeta = 8$ toward $\rmax$ at $0\degree$, i.e., $\max(\tau_*)$.

\subsubsection{Performance} \label{sec:performance}

\begin{figure*}
    \includegraphics[width=\textwidth]{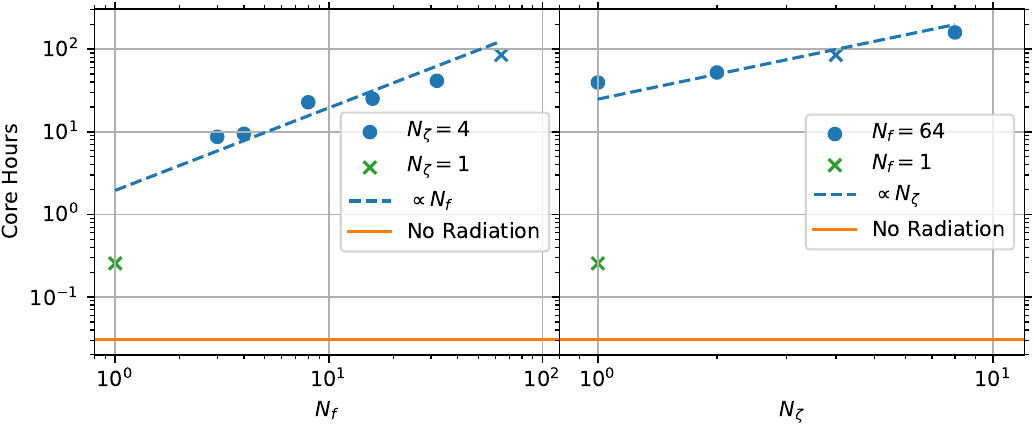}
    \caption{Total core hours (i.e., the multiplicative product of the number of CPU cores and the elapsed real time) used to integrate various Athena++ models up to the same simulation time limit $\tlim \approx 6.5\times10^{5}\s \approx 2.6\teq$ (Figure~\ref{fig:evolution} and Section~\ref{sec:performance}).
    Left and right panels are functions of the number of frequency bands $N_f$ and number of cell-centered polar angles $N_\zeta$ between 0 and $\pi/2$, respectively (Section~\ref{sec:radiation_transport}).
    Marker colors show models with different $N_\zeta$ on the left and $N_f$ on the right, crosses show the same model in either panel, dashed blue guidelines show linear scaling, and the orange horizontal line shows a similar model a without radiation.}
    \label{fig:performance}
\end{figure*}

To evaluate the performance of the Athena++ multigroup radiation transport module (Section~\ref{sec:radiation_transport}), we compare the computational costs of the additional hydrostatic models we produced to study convergence in Sections~\ref{sec:frequency_bands} and \ref{sec:angles}. 
Figure~\ref{fig:performance} and Table~\ref{tab:resolution_studies} show the total core hours (i.e., the multiplicative product of the number of CPU cores and the elapsed real time) used to integrate each model up to the same simulation time limit $\tlim$.
As a baseline, we ran an additional model with $N_f = N_\zeta = 1$ (Figure~\ref{fig:nhat}), shown by the green cross in either panel.
In general, the total core hours scale linearly with $N_f$ or $N_\zeta$ while the other parameter is kept fixed.
This is consistent with the total number of simultaneous discretized equations required by equation~\eqref{eq:radiation_transport} scaling linearly with $N_\zeta$ \citep[][\S~3.2]{Jiang2021} and $N_f$ \citep[][\S~3.2]{Jiang2022}.

As the module implicitly solves the radiative transfer equation (Section~\ref{sec:radiation_transport}), the computational performance can vary during an integration as the time-dependent solution to the radiation field evolves, e.g., as a function of $|\partial\Ertot/\partial t|$ as seen in Figure~\ref{fig:evolution}.
While $\max(\partial\log\Ertot/\partial\log t) \approx 1$ before $\tcross$, the iterative solver can require hundreds of iterations per time step to reach a total relative error threshold of $10^{-6}$ between iterative solutions \citep[][eq.~34]{Jiang2022}.
However, after $\tcross$, as the hydrostatic system begins to relax toward thermal equilibrium $\Teq$ (Section~\ref{sec:thermal_equilibrium}) and $\partial\log\Ertot/\partial\log t \ll 1$, only tens of iterations are required per time step to satisfy the same tolerance criterion.
Finally, only one iteration is needed once $\Teq$ is reached after $\teq \approx 2.5\times10^5\s$, since $\partial\Ertot/\partial t \approx 0$.\footnote{
We determined the value of $\teq$ only after running the model (i.e., a posteriori), either roughly by judging Figure~\ref{fig:evolution} or more precisely by reading the outputs that report the number of iterations between time steps.}
Although convergence for the iterative scheme can be problem dependent \citep[][\S~4]{Jiang2021}, which can include the specific opacities used (Section~\ref{sec:dust_opacities}), the total cost of each iterative step also scales linearly with $N_\zeta$ \citep[][\S~3.2.2]{Jiang2021} and $N_f$ \citep[][\S~3.2]{Jiang2022}.

To compute thermal equilibrium solutions in RADMC-3D (Section~\ref{sec:radmc-3d}), we used an average of 28.6 core hours per model (Sections~\ref{sec:radmc-3d}, \ref{sec:gray_with_dust_emission}, \ref{sec:with_dust_emission}, \ref{sec:scattering}).
We use shared-memory multiprocessing via OpenMP to parallelize the time-independent Monte Carlo radiative transfer calculations by tracing $8\times10^8$ photons per model across multiple threads.\footnote{
This differs from \textit{distributed}-memory multiprocessing, e.g., via the Message Passing Interface (MPI), which can parallelize workloads across multiple nodes.
Athena++ supports OpenMP, MPI, and hybrid parallelization \citep[][\S~2.4]{StoneTomidaWhite2020}.}
Although OpenMP alone is limited by the number of logical threads available on a single machine (or node in a cluster),
RADMC-3D shows strong scaling for the same total number of photon packets (Footnote~\ref{foot:strong_scaling}).
We emphasize that the nature of computing solutions using either code is fundamentally different:
RADMC-3D does not evolve models in time but only solves for the thermal equilibrium;
Athena++, on the other hand, evolves the radiation field in time (Sections~\ref{sec:radiation_transport} and \ref{sec:thermal_equilibrium}), and we report the steady state (i.e., thermal equilibrium) reached by the system after some thermal relaxation timescale.
In its Monte Carlo approach, RADMC-3D can use additional computational time to trace more photons, which only reduces the inherent noise in the solution.
On the other hand, Athena++ can use additional computational time to further evolve the entire radiation and temperature fields in simulated time.
As such, we note that the core hours used by Athena++ shown in Figure~\ref{fig:performance} and Table~\ref{tab:resolution_studies} result from integrating each model to $\tlim \approx 6.5\times10^5\s \approx 2.6\teq$ (Figure~\ref{fig:evolution}), as $\teq$ is a priori unknown.
Thus, the actual core hours needed by Athena++ to evolve the models only to $\teq$ is less than those reported.

To very roughly judge the added cost for Athena++ to implicitly solve the transfer equation~\eqref{eq:radiation_transport} in our irradiated disk models, we ran a similar model without radiation transport (Section~\ref{sec:radiation_transport} and Footnote~\ref{foot:implicit_flag}).
Specifically, we ran the \texttt{disk.cpp} problem generator \citep{ZhuDongStone2015} without evolving the dynamics (Section~\ref{sec:thermal_equilibrium}) but with the same domain, resolution, and disk parameters (Section~\ref{sec:stellar-irradiated_disk}).
Compared to the model with $N_f = N_\zeta = 1$, which took 619 time steps and 0.26 core hours to reach $\tlim$, this more basic run took 608 time steps and 0.03 core hours, as shown in Figure~\ref{fig:performance}.
We caution that this comparison is only a weak proxy for the additional expense to a typical hydrodynamic time step for our specific (hydrostatic) model, as the difference in core hours is disproportionately dominated by the substantial evolution of the radiation field while $t < \tcross$.
Thus, we reemphasize that the actual cost can vary widely in practice and will significantly depend on the model and choice of radiation parameters, including iterative convergence criteria (Section~\ref{sec:optimization}).

\subsection{Optimization} \label{sec:optimization}

By understanding the convergence and performance properties of our Athena++ irradiated disk models (Section~\ref{sec:resolution_studies}), we can optimize the model for the future goal of extending integrations with radiation \textit{hydrodynamics} fully enabled (cf. Section~\ref{sec:hydrostatic_comparisons}).
Keeping the number of cell-centered polar angles $N_\zeta = 4$ (cf. Figure~\ref{fig:nhat}, Section~\ref{sec:angles}, and the right panel of Figure~\ref{fig:performance}), we try to minimize the maximum absolute relative difference in equilibrium dust temperatures $\Teq$ (Section~\ref{sec:thermal_equilibrium}) with a comparative RADMC-3D model (Section~\ref{sec:radmc-3d}) while also minimizing the number $N_f$ of \textit{custom} frequency bands.
Instead of specifying only $\nu_1$ and $\nu_{N_f-1}$ to define $N_f - 2$ logarithmically uniform bands in between (Section~\ref{sec:frequency_bands}), we extend the problem generator to optionally read a separate input file specifying arbitrary band interfaces (Appendix~\ref{appx:public_repository}) which are then set in the radiation module (Section~\ref{sec:radiation_transport}).

\begin{figure*}
    \includegraphics[width=\textwidth]{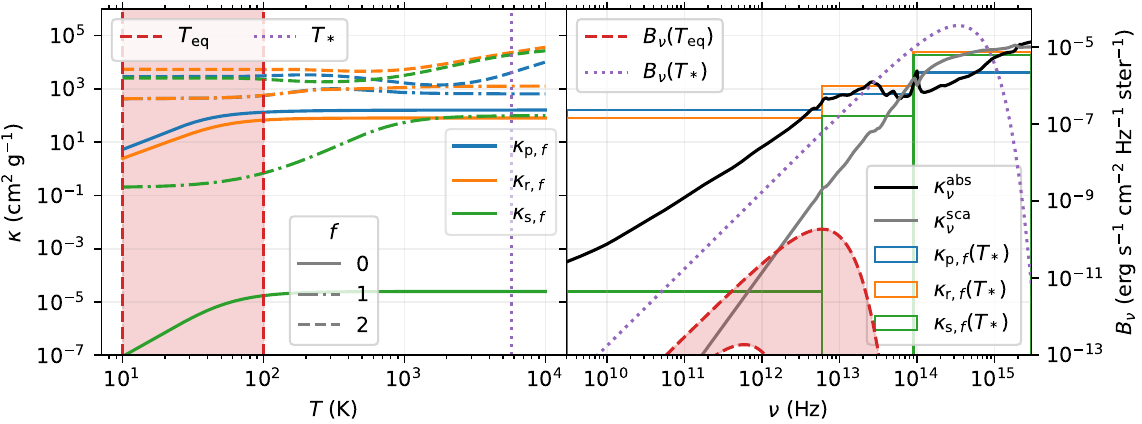}
    \caption{Similar to Figure~\ref{fig:opacities} except for only $N_f = 3$ custom, optimized frequency bands (Section~\ref{sec:optimization}).}
    \label{fig:optimized_opacities}
\end{figure*}

Figure~\ref{fig:optimized_opacities} shows the mean opacities for $N_f = 3$ custom bands as an example of an optimized model.
We choose $\nu_1 = 6\times10^{12}$ and $\nu_2 = 9\times10^{13}$ to best capture the relatively flat profile of $\knuabs$ within the $f = 1$ band and also take advantage of the sharp bump in $\knuabs$ at $10^{14}\Hz$ to maximize the mean absorption opacities (e.g., $\kpf$) for the highest $f = 2$ band.
As explained in Section~\ref{sec:frequency_bands}, maximizing absorption in bands containing $\nu \gtrapprox \nu_*$ helps produce more accurate $\Teq$ (Figure~\ref{fig:nfreq}, lower panels) in the effectively thin disk atmosphere (Figure~\ref{fig:disk}), since $\log\knuabs \propto \log\nu$.

\begin{figure*}
    \includegraphics[width=\textwidth]{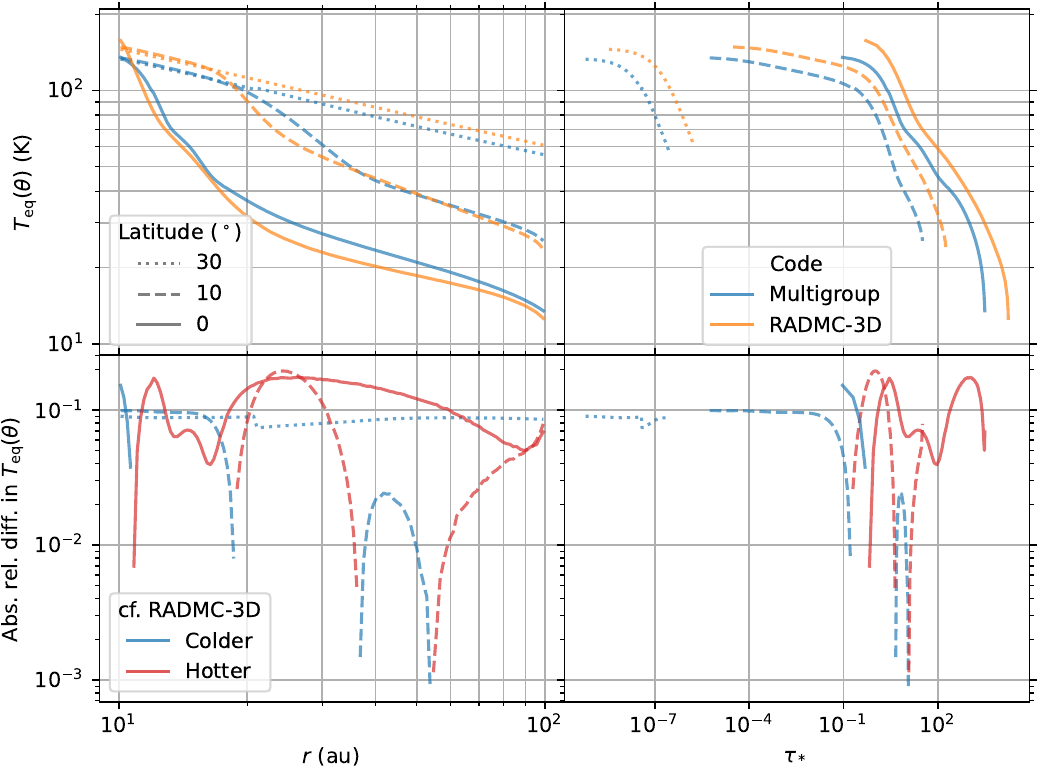}
    \caption{Similar to Figure~\ref{fig:scattering} except for an optimized Athena++ model with $N_f = 3$ custom frequency bands (Section~\ref{sec:optimization}) and the corresponding band-mean opacities shown in Figure~\ref{fig:optimized_opacities}.}
    \label{fig:optimized_profiles}
\end{figure*}

Figure~\ref{fig:optimized_profiles} and Table~\ref{tab:resolution_studies} compares this optimized model with RADMC-3D.
As a function of the effective depth to peak stellar irradiation $\tau_*$ [equation~\eqref{eq:tau_*}] in the lower-right panel, the absolute relative difference in $\Teq$ is at most 9\% at latitudes of $30\degree$ and $10\degree$ for $\tau_* < 1$, while remaining cooler (cf. Sections~\ref{sec:with_dust_emission} and \ref{sec:scattering}).
Compared to the $N_f = 64$ model (Section~\ref{sec:scattering}), the absolute relative difference in $\Teq$ with RADMC-3D is 17\% (or a factor of two greater) along the midplane (cf. Figure~\ref{fig:scattering}), but this model with $N_f = 3$ takes at least a factor of ten fewer core hours to integrate up to the same simulation time limit (Figure~\ref{fig:performance}, left panel).
Thus, accuracy and performance can be exchanged depending on the desired application or production constraints.

To further optimize performance, we recommend reducing the implicit speed of light $c$ in equation~\eqref{eq:radiation_transport} as needed by setting the optional parameter \texttt{reduced\_factor} to a fraction of unity under the \texttt{<radiation>} block in the input file (Appendix~\ref{appx:public_repository}).\footnote{
This can be done (or undone) before starting a simulation or between restarts.}
Section~\ref{sec:performance} explains how substantial changes in the radiation field between hydrodynamic time steps can significantly increase the number of iterative steps needed by the implicit solver \citep[][\S~3.2]{Jiang2022}.
Limiting the speed of light reduces the (spatial) extent to which the (local) radiation field can change, and thus the number of iterations needed, between time steps.
To significantly reduce the number of iterations needed while $t < \tcross$ (Figure~\ref{fig:evolution}), we use $0.01c$ for all Athena++ models presented here, which remains at least $10^4$ faster than the locally isothermal sound speed $\csiso$ [cf. equation~\eqref{eq:csiso}].

Reducing the speed of light does not change our hydrostatic solutions, but doing so can affect the time-dependent evolution of hydrodynamic models (Section~\ref{sec:applications}).
Although it can decrease the number of iterations needed between time steps, it typically also increases the overall time required to reach a steady state.
As an alternative approach, such irradiated models can first be hydrostatically evolved to thermal equilibrium (as we do here, following Section~\ref{sec:thermal_equilibrium}, with or without a reduced speed of light), before restarting with hydrodynamics enabled and evolving them to a steady state.

\section{Discussion} \label{sec:discussion}

This section discusses our multigroup radiation transport framework (Section~\ref{sec:methodology}) and our stellar-irradiated disk model (Section~\ref{sec:stellar-irradiated_disk}) in the broader context of protoplanetary disk modeling.
Section~\ref{sec:alternative-approaches} contrasts this framework with alternative radiation-hydrodynamic approaches and implementations.
Section~\ref{sec:applications} outlines various applications of our framework to self-consistently investigate the hydrodynamic and thermodynamic properties of stellar-irradiated disks.

\subsection{Alternative Approaches} \label{sec:alternative-approaches}

In an attempt to account for the frequency dependence of stellar radiation, \cite{KuiperKlahrDullemond2010} combined flux-limited diffusion (Section~\ref{sec:introduction}) with ray tracing in the magnetohydrodynamics code PLUTO \citep{MignoneBodoMassaglia2007}.
Specifically, stellar irradiation was modeled with a first-order ray-tracing routine with frequency-dependent absorption opacities [cf. equation~\eqref{eq:F_r-tau}], while dust thermal reemission was approximated by flux-limited diffusion with gray opacity.
However, scattering is neglected, as it makes the radiative transfer equation integrodifferential \citep[][\S~1.7]{RybickiLightman1979}.

Similar to Section~\ref{sec:with_dust_emission}, \citet[][\S~3.3.3; Figs.~5, 6]{KuiperKlahrDullemond2010} compared the temperature profiles of their static models with those from the predecessor of RADMC-3D (Section~\ref{sec:radmc-3d}), RADMC \citep{DullemondTurolla2000, DullemondDominik2004}.
Along an optically thick disk midplane, they reported relative errors of up to 11.1\% for their hybrid method with frequency-dependent irradiation but up to 38.4\% with gray irradiation.
They also showed qualitative agreement of meridional (i.e., polar) profiles between their hybrid method with frequency-dependent irradiation and RADMC but only at $r = 2\au$.
However, in an extended follow-up comparison of their frequency-dependent hybrid method and RADMC, \citet[][\S~5.2.4; Tab.~1; Figs.~7, 8]{KuiperKlessen2013} found temperatures along an even more optically thick midplane deviated as much as 48\% as $\tau$ approaches $10^3$ and remained above 35\% for $10^3 \lesssim \tau \lesssim 10^6$ (cf. Table~\ref{tab:hydrostatic_comparisons} and Figure~\ref{fig:absorption}).

Following \cite{KuiperKlahrDullemond2010}, \cite{KriegerKlahrMelonFuksman2025} implemented this hybrid method into the hydrodynamic code TRAMP \citep{KlahrHenningKley1999}.
However, similar to \cite{ZhangZhuJiang2024}, they used only mean (i.e., gray) opacities for both the dust thermal reemission and the ray-traced stellar irradiation, each evaluated at the respective local dust and effective stellar temperatures \citep[][eqs.~3 and 4, respectively]{KriegerKlahrMelonFuksman2025}, instead of frequency-dependent opacities and flux for the latter.
Although they assumed a global dust-to-gas mass ratio of 1\% in TRAMP, they used the grid cell temperatures resulting from their radiation-hydrodynamic snapshots to adjust the local dust densities \citep[][eq.~12]{KriegerKlahrMelonFuksman2025} in their comparative hydrostatic models produced with the Monte Carlo radiative transfer code Mol3D \citep{OberWolfUribe2015}.
Despite this a posteriori modification to improve the consistency of absorption coefficients \citep[i.e., $\rho\kappa$;][eq.~1.22]{RybickiLightman1979} between models, they found relative differences as high as 33\% \citep[][\S~3.1.1 and Figs.~3, 4]{KriegerKlahrMelonFuksman2025}.

In a recent attempt to address the artificial interactions between crossing beams inherent to the M1 closure method (Section~\ref{sec:introduction}), \cite{MelonFuksmanFlockKlahr2025} implemented into PLUTO a half-moment closure method \citep{DubrocaKlar2002}, which integrates the specific intensity over hemispheres instead of all solid angles.
Neglecting scattering, their M1 and half-moment disk models with 22 frequency groups disagree with RADMC-3D by up to 21\% and 6\%, respectively, or with 3 groups by up to 23\% and 8\%, respectively.
However, the half-moment method cannot be generalized to all radiative interactions, as allowable beam crossings are restricted only to a preselected direction \citep[][Fig.~2]{MelonFuksmanFlockKlahr2025}.

\subsection{Potential Applications} \label{sec:applications}

Using the larger Athena++ hydrodynamic framework (Section~\ref{sec:methodology}), our multigroup framework can facilitate radiation hydrodynamics investigations into a variety of protoplanetary disk models with frequency-dependent stellar irradiation, absorption, and scattering.
To compare with hydrostatic models from Monte Carlo radiative transfer methods, we set $\rho_0$ (Section~\ref{sec:stellar-irradiated_disk}) so the density field defined by equation~\eqref{eq:rho} represents only dust with the size distribution specified in Section~\ref{sec:dust_opacities}.
However, depending on the dust-to-gas mass ratio, $\rho_0$ and the dust opacities can be adjusted when evolving the gas disk (cf. Section~\ref{sec:thermal_equilibrium}), which RADMC-3D cannot do.

Our current model (Section~\ref{sec:stellar-irradiated_disk}) straightforwardly applies to disks with large-scale substructures, e.g., depleted gaps.
Although it was designed to mitigate challenges to the implicit solver from an optically thick inner radial boundary to stellar irradiation, our radially softened density profile, defined by equation~\eqref{eq:rhosig} and shown in Figure~\ref{fig:midplane} (blue curve), can effectively model a disk outside the orbit of a planet that carved a gap at or within $10\au$. 
In this case, equations~\eqref{eq:I_f-rout} and \eqref{eq:I-rout} could be adjusted to more accurately capture the prior attenuation of stellar radiation by dust inside the orbit (i.e., $r < \rmin$).

On the other hand, since equations~\eqref{eq:I_f-rout} and \eqref{eq:I-rout} assume $R_* < r < \rmin$ is optically thin, our current flux of stellar irradiation through the radial boundary readily models a transitional disk \citep{EspaillatMuzerolleNajita2014} with an inner dust-depleted cavity \citep{VioqueKurtovicTrapman2025}.
Despite having truncated dust distributions, transitional sources exhibit active accretion, indicating the presence of gas in the cavity.
In the  hydrodynamic case, this dust truncation radius, including this optically thin inner region of gas, can be modeled by softening the opacity field, instead of the (gas) density as in equation~\eqref{eq:rhosig}.
As suggested in Section~\ref{sec:stellar-irradiated_disk}, softening the opacity coefficients [equations~\eqref{eq:kpfe}--\eqref{eq:ksf}] over some radial length can alternatively maintain the effectively thin boundary to peak stellar irradiation $\tau_*$ [equation~\eqref{eq:tau_*}] shown in Figure~\ref{fig:midplane} (green curve) while allowing the gas density profile to follow a simple power law \citep[e.g.,][eq.~3]{NelsonGresselUmurhan2013}.

Since Athena++ evolves models in time (Section~\ref{sec:thermal_equilibrium}; Figure~\ref{fig:evolution}), unlike RADMC-3D, the variable luminosity from accretion, as observed for transitional disks and protoplanetary disks in general, can also motivate applications of our framework to dynamic irradiation sources.
For example, sporadic yet intense frequency-dependent radiation from stellar outbursts can be modeled with time-dependent versions of equations~\eqref{eq:I_f-rout} or \eqref{eq:I-rout}.
Similarly, stellar irradiated circumbinary disks can be studied by making these equations periodic.

Our framework can also be used to study disk chemistry (Section~\ref{sec:introduction}).
Frequency-dependent radiative transfer has previously been combined with various (photo)chemical and physical processes to compute hydrostatic disk models and interpret gas emission lines \citep{WoitkeKampThi2009}.
Moreover, established thermochemical codes already use Monte Carlo methods \citep{DuBergin2014} to model the chemical evolution \citep{AlarconTeagueZhang2020} and thermal structures of observed disks \citep{CalahanBerginZhang2021}.
Thus, the flexibility of our framework could complement or even be combined with thermochemical codes to simultaneously evolve the distribution, phases, and interactions of molecules relevant to planet formation.
As disk dynamics affects temperature, which affects chemistry, which affects ionization, which affects disk dynamics, this coupled feedback loop should be studied self-consistently, which cannot be done in RADMC-3D.

\section{Conclusions} \label{sec:conclusions}

We introduce a comprehensive framework for developing accurate, self-consistent irradiated disk models with frequency-dependent absorption and scattering for radiation hydrodynamics in Athena++ (Section~\ref{sec:methodology}).
We added radial rays to the multigroup radiation transport module (Section~\ref{sec:radiation_transport}) to capture stellar irradiation (Section~\ref{sec:stellar_irradiation} and Figure~\ref{fig:nhat}) and systematically evaluate band-mean opacities (Section~\ref{sec:emissivity_opacity} and Figure~\ref{fig:opacities}) at the local or color temperature depending on irradiative--thermal equilibrium (Section~\ref{sec:irradiative--thermal_equilibrium}).
To demonstrate our framework in the hydrostatic limit (Section~\ref{sec:model}), we formulate an axisymmetric disk model (Figures~\ref{fig:midplane}, \ref{fig:vertical}, and \ref{fig:disk}), irradiated by a sunlike star (Section~\ref{sec:stellar-irradiated_disk}); 
and using the dust opacities adopted by the DSHARP project (Section~\ref{sec:dust_opacities}), we hydrostatically evolve the radiation field (Figure~\ref{fig:evolution}) until it reaches thermal equilibrium (Section~\ref{sec:thermal_equilibrium} and Figure~\ref{fig:thermal_equilibrium}).

We thoroughly study the accuracy of our framework by comparing the equilibrium temperatures of hydrostatic models (Section~\ref{sec:hydrostatic_comparisons}) primarily against the Monte Carlo radiative transfer code RADMC-3D (Section~\ref{sec:radmc-3d}).
Compared to an analytic solution for purely attenuated ray tracing (Section~\ref{sec:without_dust_emission}), the radial flux is slightly higher, but the relative error decreases linearly overall with increasing radial resolution (Figure~\ref{fig:attenuation}). 
Across at least 13 orders of magnitude in optical depth that can reach above $10^4$ (Table~\ref{tab:hydrostatic_comparisons}), our models with 64 frequency bands differ from RADMC-3D (with 150 wavelength points) by less than 6\% on average and at most 11\% for gray opacity (Section~\ref{sec:gray_with_dust_emission}), but similar on average and at most 8\% for frequency-dependent absorption (Section~\ref{sec:with_dust_emission}) and with scattering (Section~\ref{sec:scattering}).
Compared to gray models, the difference between atmospheric and midplane temperatures is at least three times greater with frequency-dependent opacities (cf. Figures~\ref{fig:gray_absorption}, \ref{fig:absorption}, and Figure~\ref{fig:scattering}).
Moreover, the difference between isotropic and anisotropic scattering is less than 4\% on average and at most 9\% (Figure~\ref{fig:scattering_modes}).

We also study the convergence and performance of our framework with respect to key resolution parameters in the radiation module (Section~\ref{sec:resolution_studies}).
In general, increasing the number of frequency bands (Section~\ref{sec:frequency_bands}) or angular rays (Section~\ref{sec:angles}) increases the accuracy (Figures~\ref{fig:nfreq} and \ref{fig:nzeta}) but also linearly increases the total computational time (Section~\ref{sec:performance} and Figure~\ref{fig:performance}).
However, reducing 64 logarithmically uniform bands to just 3 optimized bands (Figure~\ref{fig:optimized_opacities}) reduces the computation time by at least an order of magnitude while only increasing potential inaccuracies by two to three times (Section~\ref{sec:optimization} and Figure~\ref{fig:optimized_profiles}).

We discuss our framework in the broader context of protoplanetary disk modeling (Section~\ref{sec:discussion}).
We review the limitations of alternative approaches to radiation hydrodynamics (Section~\ref{sec:alternative-approaches}), including hybrid methods using frequency-dependent ray-tracing and flux-limited diffusion (cf. Section~\ref{sec:hybrid_ray_tracing}) as well as a half-moment variation of the M1 closure.
For the former, ray tracing is unable to incorporate scattering effects, and comparisons with Monte Carlo methods showed deviations as high as 48\% with increasing optical depth.
Since radiative energy scales with $T^4$, such temperature deviations can produce substantial errors in the estimated energy budget, reinforcing the need to accurately capture radiative heating.
For the latter, the half-moment closure allows radiation beams to cross but only in one preselected direction.

Lastly, we outline potential applications of our framework to examine the hydrodynamic properties of stellar-irradiated disks (Section~\ref{sec:applications}).
Our current model already applies to transitional disks or can be tuned for those with large-scale substructures, e.g., planet-carved gaps.
As the Athena++ hydrodynamic code evolves systems in time, our flexible framework can model dynamic irradiation (e.g., from accretion outbursts or binary host stars) or compliment thermochemical codes.
We maintain and make available our open-source resources in a public repository (Appendix~\ref{appx:public_repository}).

\begin{acknowledgments}
SAB acknowledges support from the Simons Foundation's Flatiron Institute; its Center for Computational Astrophysics Pre-Doctoral Program, during which this work was initiated; its Scientific Computing Core, which facilitated our computations; and funding from the UNLV Foundation Board of Trustees Fellowship.
SAB also acknowledges Andrew Youdin, Wladimir Lyra, and the Planet Formation in the Southwest Plus (PFITS+) Collaboration for helpful discussions.
ZZ acknowledges support from NASA award 80NSSC22K1413 and 80NSSC25K7144 and NSF award 2429732 and 2408207.
SZ acknowledges support by NASA through the NASA Hubble Fellowship grant \#HST-HF2-51568 awarded by the Space Telescope Science Institute, which is operated by the
Association of Universities for Research in Astronomy, Inc., for NASA, under contract NAS5-26555.
\end{acknowledgments}

\begin{contribution}



SAB wrote and revised the manuscript; managed and coordinated research planning and execution; implemented, tested, and helped extend the Athena++ radiation transport module; developed the methodology (including mean-opacity scripts) and models (including RADMC-3D and Athena++ problem generators); produced the data and figures; collected, investigated, analyzed, and validated the data; and created and maintains the public project repository.
All coauthors reviewed and commented on the manuscript.
Y-FJ acquired funding for and supported the supervision of the project, supported the use and development of the code, committed code changes to the public version of Athena++, and provided high-performance computing resources.
ZZ suggested the original research concepts, guided evolving objectives, critically reviewed the manuscript, and provided helpful suggestions for its development and improvement.
SZ wrote Appendix~\ref{appx:hybrid_ray-tracing_method}, developed and produced the hybrid ray-tracing models, calculated and provided the opacities, independently reproduced the models, and validated the data.
PJA consulted on, and supported the development and supervision of, the project.

\end{contribution}

\software{
Athena++ \citep{StoneTomidaWhite2020}, DSHARP Mie-Opacity Library \citep{BirnstielDullemondZhu2018}, Matplotlib \citep{Hunter2007}, NumPy \citep{HarrisMillmanvanderWalt2020}, RADMC-3D \citep{DullemondJuhaszPohl2012}, SciPy \citep{VirtanenGommersOliphant2020}
          }


\appendix

\section{Public Repository} \label{appx:public_repository}

To make this framework accessible, we have created a public GitHub repository\footnote{ \url{https://github.com/sabaronett/irrad_disk}}
\citep{Baronett2025} to maintain the relevant resources.
These include our problem generator (which may also be included in a future public version of Athena++), described in Sections~\ref{sec:stellar_irradiation}, \ref{sec:irradiative--thermal_equilibrium}, and \ref{sec:stellar-irradiated_disk}; the modified Athena++ source files needed to prevent the disk from dynamically evolving (Section~\ref{sec:thermal_equilibrium});
input files for every model presented in Section~\ref{sec:hydrostatic_comparisons};
the tabulated monochromatic dust opacities we used (Section~\ref{sec:dust_opacities});
separate Python scripts to precompute band-mean opacity coefficients based on an Athena++ input file (Section~\ref{sec:emissivity_opacity}) and to create and format the required input files for each model presented in Section~\ref{sec:hydrostatic_comparisons};
and Jupyter Notebooks to generate all of the figures.
We invite users to create a GitHub issue in the repository for any questions related to these items, problems encountered in their use, or to provide feedback.







\section{Modified Angular Discretization} \label{appx:modified_angular_discretization}

The angular discretization implemented in Athena++ by \cite{Jiang2021} was first introduced into Athena \citep{StoneGardinerTeuben2008} by \citet[][\S~3.2]{DavisStoneJiang2012}, according to the algorithm described by \citet[][Appx.~B]{BrulsVollmollerSchussler1999}, based on the principles of type-A angular quadratures described by \cite{Carlson1963}.
Continuing from Section~\ref{sec:radiation_transport}, each angular unit vector is uniquely specified by three components $\nhat = (\mu_x, \mu_y, \mu_z)$, corresponding to its direction cosines \citep[][\S~3]{JiangStoneDavis2014}.
If \texttt{angle\_flag = 1}, \citep[cf.][eq.~6]{DavisGammie2020}
\begin{align}
    \mu_x &= \sin\zeta\cos\psi, \\
    \mu_y &= \sin\zeta\sin\psi, \\
    \mu_z &= \cos\zeta,
\end{align}
where local longitudes $\psi \in (0, 2\pi)$ and colatitudes (i.e., polar angles) in $\cos\zeta$ space for $\zeta \in [0, \pi]$ are both evenly divided \citep[][\S~3.2.4]{Jiang2021}.
The directions of $\nhat$ must ensure the $\Nang$ solid angles $\Omega$ they subtend mutually cover the entire unit sphere while being equal in weight $w$ 
\citep[cf.][Appx.~B]{BrulsVollmollerSchussler1999} to satisfy \citep[cf.][eq.~11]{Jiang2021}
\begin{align}
    \label{eq:0th-moment_angles}
    \sum_{n=0}^{\Nang-1}w_n &= 1, \\
    \label{eq:1st-moment_angles}
    \sum_{n=0}^{\Nang-1}\mu_{x,n}w_n &= \sum_{n=0}^{\Nang-1}\mu_{y,n}w_n = \sum_{n=0}^{\Nang-1}\mu_{z,n}w_n = 0, \\
    \label{eq:2nd-moment_angles}
    \sum_{n=0}^{\Nang-1}\mu_{x,n}^2 w_n &= \sum_{n=0}^{\Nang-1}\mu_{y,n}^2 w_n = \sum_{n=0}^{\Nang-1}\mu_{z,n}^2 w_n = \frac{1}{3}.
\end{align}

By including the new pair of $\nhat_r$ via \texttt{polar\_angle = 1} (Section~\ref{sec:stellar_irradiation}), we must slightly modify the original discretization algorithm to satisfy equations~\eqref{eq:0th-moment_angles}--\eqref{eq:2nd-moment_angles}.
Since we choose to exclude $\nhat_r$ from $N_\zeta$,
\begin{equation}
    \Nang = 4N_\zeta N_\psi + 2.
\end{equation}
This does not affect equation~\eqref{eq:0th-moment_angles}, since $w_n = 1/\Nang$ by definition (Section~\ref{sec:radiation_transport}).
Moreover, because $\mu_z = \pm 1$ and $\mu_x = \mu_y = 0$ for $\nhat_r$, $\mu_x$ and $\mu_y$ for nonpolar $\nhat$ remain the same.
However, we must redetermine $\mu_z$ for these other $\nhat$ to first satisfy $\sum_{n=0}^{\Nang-1}\mu_{z,n}w_n = 0$ in equation~\eqref{eq:1st-moment_angles}.

We focus our analysis on the spherical wedge $\psi \in [0, \pi/N_\psi]$, as the results in $\zeta$ can be extended to the remaining wedges via $2N_\psi - 1$ rotations by $\pi/N_\psi$ in $\psi$.
There are $2N_\zeta + 1$ colatitudes (i.e., interfaces) $\zetaf_i$, $i \in \mathbb{N}^0$, that latitudinally bound the solid angles $\Omega$.
By reflection about $\zetaf_{N_\zeta} = \pi/2$ at the equator,
\begin{equation}
    \cos\zetaf_i = -\cos\zetaf_{2N_\zeta-i},
    \label{eq:reflection}
\end{equation}
since $\cos(\pi - \zeta) = -\cos\zeta$.
For the $2N_\zeta$ midlatitude solid angles $\Omegamid$ in the wedge,
\begin{align}
    \Omegamid &= \int_{\zetaf_i}^{\zetaf_{i+1}}\sin\zeta\od\zeta\int_0^{\pi/N_\psi}\od\psi \\
    \label{eq:Omegamid}
        &= \frac{\pi}{N_\psi}\left(\cos\zetaf_i - \cos\zetaf_{i+1}\right).
\end{align}
If $\zetaf_0$ bounds the polar solid angle $\Omegapol$ subtended by $\nhatrout$ to the north, then
\begin{align}
    \Omegapol &= \int_0^{\zetaf_0}\sin\zeta\od\zeta\int_0^{2\pi}\od\psi \\
    \label{eq:Omegapol}
        &= 2\pi\left(1 - \cos\zetaf_0\right),
\end{align}
where we note $\Omegapol$ spans a spherical cap, hence $\int_0^{2\pi}\od\psi$.
At the south pole, $\nhatrin$ also subtends $\Omegapol$, as can be verified by evaluating $\int_{\zetaf_{2N_\zeta}}^\pi\sin\zeta\od\zeta$ using equation~\eqref{eq:reflection}.

Since all solid angles $\Omega$ must be equal in magnitude,
\begin{align}
    \Nang\Omega &= 4N_\zeta N_\psi\Omegamid + 2\Omegapol \\
        &= 2\pi\sum_{i = 0}^{2N_\zeta - 1}\left(\cos\zetaf_i - \cos\zetaf_{i+1}\right) \notag \\
            &\quad + 4\pi(1 - \cos\zetaf_0).
\end{align}
Using equation~\eqref{eq:reflection} to cancel all expanded $\cos\zetaf_i$ terms yields
\begin{equation}
    \Omega = \frac{4\pi}{\Nang}.
    \label{eq:Omega}
\end{equation}
Combining equations~\eqref{eq:Omegapol} and \eqref{eq:Omega} yields
\begin{equation}
    \cos\zetaf_0 = \frac{\Nang - 2}{\Nang}.
\end{equation}
Since nonpolar $\nhat$ must be evenly divided in $\cos\zeta$,
\begin{equation}
    \Delta\cos\zeta \equiv \frac{\cos\zetaf_0}{N_\zeta},
    \label{eq:coszetaf_i}
\end{equation}
which can be verified by combining equations~\eqref{eq:Omegamid} and \eqref{eq:Omega} with $\Delta\cos\zeta \equiv \cos\zetaf_i - \cos\zetaf_{i+1}$.
Thus,
\begin{equation}
    \cos\zetaf_i = \cos\zetaf_0 - i\Delta\cos\zeta,
\end{equation}
for all latitudinal interfaces, and
\begin{equation}
    \cos\zeta_i^\prime = \frac{\cos\zetaf_i - \cos\zetaf_{i+1}}{2},\quad i < 2N_\zeta.
    \label{eq:mu_zi-prime}
\end{equation}

Although equation~\eqref{eq:mu_zi-prime} provisionally satisfies the $\mu_{z,n}$ criterion for the first moment of $w_n$ in equation~\eqref{eq:1st-moment_angles}, we introduce the renormalization factor $b$ to satisfy that for the second moment.
By extracting $\mu_z^2 = 1$ for each $\nhat_r$ from equation~\eqref{eq:2nd-moment_angles},
\begin{align}
    \frac{1}{3w_n} &= \sum_{n=0}^{\Nang - 1}\mu_{z,n}^2 \\
    \frac{\Nang}{3} &= 2 + 2N_\psi\sum_{i = 0}^{2N_\zeta - 1}\left(b\cos\zeta_i^\prime\right)^2 \\
    b &= \left(\frac{\Nang - 6}{6N_\psi\sum_{i = 0}^{2N_\zeta - 1}{\cos\zeta_i^\prime}^2}\right)^{1/2},
\end{align}
which can be evaluated numerically.
Finally,
\begin{equation}
    \cos\zeta_i = b\cos\zeta_i^\prime,\quad i < 2N_\zeta,
\end{equation}
for all nonpolar $\nhat$.
Compared to the originals in Figure~\ref{fig:nhat}, these new $\mu_{z,n}$ imply a uniformly slight deflection of blue $\nhat$ toward the local equator and away from (i.e., to accommodate $\Omegapol$ subtended by) the orange $\nhat_r$ at the poles.

\section{Hybrid Ray-Tracing Method} \label{appx:hybrid_ray-tracing_method}

The hybrid ray-tracing scheme computes the stellar irradiation as the sum of wavelength-dependent fluxes, whereas the disk thermal radiation is treated in the single-frequency approximation, following the implementation of \cite{KuiperKlahrDullemond2010}. 
Compared to the two-temperature approach (one temperature for the star and one for the disk) used in \cite{ZhangZhuJiang2024}, this method improves the accuracy of the temperature calculations.
This is because longer-wavelength stellar photons, which exhibit lower optical depths, can penetrate deeper into the disk and heat regions closer to the midplane.
In other words, the surface where $\tau = 1$ is no longer a sharp boundary. 
Instead, each frequency has its own surface, where $\tau_\nu = 1$, and longer wavelength infrared photons have $\tau = 1$ surfaces closer to the midplane.

To calculate the heating from stellar irradiation, we compute the radial flux in each frequency bin as [cf. equation~\eqref{eq:F_r-tau}]:
\begin{equation}
F_*(\nu_i) = e^{-\tau_{\nu_i}^\mathrm{abs}}  \frac{c}{4} \int_{(\nu_{i-1} + \nu_i)/2}^{(\nu_i + \nu_{i+1})/2} B_\nu(T_*)\od\nu,
\end{equation}
where the optical depth to absorption at frequency $\nu$ is given by [cf. equation~\eqref{eq:tau_*}]:
\begin{equation}
\tau_\nu^\mathrm{abs}(r, \theta, \phi) = \int_{\rmin}^{r} \rho(r^\prime, \theta, \phi)\knuabs \od r^\prime.
\end{equation}
The total stellar irradiation flux is then the sum over all frequency bins:
\begin{equation}
F_{*,\mathrm{tot}} = \sum_i F_*(\nu_i).
\end{equation}
If the opacity is wavelength independent, this expression reduces to equation~\eqref{eq:F_r-tau}.
We use the same monochromatic absorption opacity specified in Section \ref{sec:dust_opacities} with 200 logarithmically uniform wavelength bins, sampling between 0.1 $\mu$m and 10 cm.

\bibliography{refs}{}
\bibliographystyle{aasjournalv7}



\end{CJK*}
\end{document}